\documentclass[twocolumn]{aa}
\usepackage[referable]{threeparttablex}
  
\def\fds{\hbox{$.\!\!{''}$}}  
  
\usepackage{amsmath}
\usepackage{newtx}
\usepackage{natbib}
\usepackage{graphicx}
\usepackage{colortbl}
\usepackage{comment}

\begin{document}
\title{The Outer Rings of SN\,1987A from year 1994 to 2024}
\subtitle{Morphology, Light Curves, and Optical to Mid-Infrared Spectra}
\titlerunning{The outer rings of SN\,1987A from 1994 to 2024}
\authorrunning{Rosu et al.}

\author{S. Rosu\inst{1} \fnmsep\corrauth{sophie.rosu@unige.ch} \fnmsep \thanks{Equally contributed to this article.} \and E. Gerville-Reache\inst{2,1}\fnmsep \thanks{Equally contributed to this article.}\email{elkogerville@gmail.com} \and S. Thomas\inst{1} \email{Steven.Thomas@etu.unige.ch} \and J. Larsson\inst{3} \email{josla@kth.se} \and P. J. Kavanagh\inst{4} \email{patrick.kavanagh@mu.ie} \and J. Spyromilio\inst{5} \email{jspyromi@eso.org} \and C. Fransson\inst{6} \email{claes@astro.su.se} \and C. Gall\inst{7} \email{christa.gall@nbi.ku.dk} \and R. D. Gehrz\inst{8} \email{gehrz001@umn.edu} \and A. S. Hirschauer\inst{9} \email{alechirschauer@gmail.com} \and O. C. Jones\inst{10} \email{olivia.jones@stfc.ac.uk} \and R. P. Kirshner\inst{11} \email{rkirshner@cfa.harvard.edu} \and P. Lundqvist\inst{6} \email{peter@astro.su.se} \and M. Matsuura\inst{12} \email{matsuuram@cardiff.ac.uk} \and M. Meixner\inst{13} \email{margaret.meixner@jpl.nasa.gov} \and B. Sargent\inst{14,15} \email{bsargent@setiap.org} \and J. Sollerman\inst{6} \email{jesper@astro.su.se}}

\institute{Department of Astronomy, University of Geneva, Chemin Pegasi 51, 1290 Versoix, Switzerland \and Leiden Observatory, University of Leiden, Einsteinweg 55, 2333 CC, Leiden, The Netherlands \and Department of Physics, KTH Royal Institute of Technology, The Oskar Klein Centre, AlbaNova, SE-106 91 Stockholm, Sweden \and Department of Physics, Maynooth University, Maynooth, Co. Kildare, Ireland \and European Southern Observatory, Karl-Schwarzschild-Str 2, Garching, D-85748, Germany \and Department of Astronomy, Stockholm University, The Oskar Klein Centre, AlbaNova, SE-106 91 Stockholm, Sweden \and DARK, Niels Bohr Institute, University of Copenhagen, Jagtvej 155A, DK-2200 Copenhagen \and Minnesota Institute for Astrophysics, School of Physics and Astronomy, University of Minnesota, 116 Church St. S.E., Minneapolis, MN 55455, USA \and Department of Physics \& Engineering Physics, Morgan State University, 1700 East Cold Spring Lane, Baltimore, MD 21251, USA \and UK Astronomy Technology Centre, Royal Observatory, Blackford Hill, Edinburgh, EH9 3HJ, UK  \and TMT International Observatory, Pasadena CA 91124 USA \and Cardiff Hub for Astrophysics Research and Technology (CHART), School of Physics and Astronomy, Cardiff University, Queen's Buildings, The Parade, Cardiff CF24 3AA, UK \and Jet Propulsion Laboratory, California Institute of Technology, 4800 Oak Grove Dr., Pasadena, CA 91109, USA \and Space Telescope Science Institute, 3700 San Martin Drive, Baltimore, MD 21218, USA \and Center for Astrophysical Sciences, The William H. Miller III Department of Physics and Astronomy, Johns Hopkins University, Baltimore, MD 21218, USA}

\date{}

\abstract{The outer rings (ORs) of Supernova (SN) 1987A were ejected some 20\,000 years before the explosion. Their characterisation is crucial for constraining the properties of the progenitor of this famous supernova.}
{While numerous studies investigated in detail the SN ejecta, equatorial ring (ER), and reverse shocks, very few were dedicated to the ORs. We fill this gap and investigate the ORs physical properties. We analyse data obtained over a long temporal period, from multiple instruments, and over a wide wavelength range from optical to mid-infrared of the northern and southern ORs (NOR and SOR) of SN\,1987A. We combine observations taken with the Hubble Space Telescope (HST) between 1994 and 2022, with VLT/MUSE in 2023, and with the James Webb Space Telescope (JWST) in 2022 and 2024.}
{We measure emission flux in different regions of the ORs in narrow-band HST filter and broad- and narrow-band JWST/NIRCam filter images. We extract optical and mid-infrared spectra for the ORs in the MUSE and JWST/MIRI/MRS data and measure line emission fluxes. We analyse the evolution of the ORs clumps' morphology over time with HST.}
{The optical light curves of the ORs have shown a steady decline with time over the last 30 years, both in the H$\alpha$ and [O\,\textsc{iii}] $\lambda$\,5007 filters. This behaviour is expected as the ORs were ionised by the initial supernova UV-flash and are since then fading. The observations do not show any sign of interaction of the SN ejecta with the ORs, that would appear as a brightening in the ORs' optical light curve. We estimated the decay times for [O\,\textsc{iii}] to be 900 and 630-730\,days for the NOR and SOR, respectively, and for H$\alpha$+[N\,\textsc{ii}] to be 15\,870 and 7160\,days for the NOR and SOR, respectively. We provide the first mid-resolution spectra of the ORs in the optical and mid-infrared. We constrained the temperature from the optical [N\,\textsc{ii}] lines to $13400-16900$\,K and $11800-14500$\,K for the NOR and SOR, respectively. We constrained the electron density from the optical [S\,\textsc{ii}] lines to $610-670$\,cm$^{-3}$ and $720-790$\,cm$^{-3}$ for the NOR and SOR, respectively. }
{The spectra of the ORs differ significantly from the spectrum of the ER, not only in the lines detected, but also in the line ratios of given lines. The ORs will likely keep on fading for the next years, until the SN ejecta sweep up the ORs. Continued monitoring of SN 1987A and its ring system in the X-ray to mid-infrared is essential to capture this instant. }


\maketitle
\nolinenumbers

\section{Introduction\label{sect:introduction}}
\begin{figure}
\includegraphics[clip=true, trim=70 60 55 60,width=\linewidth]{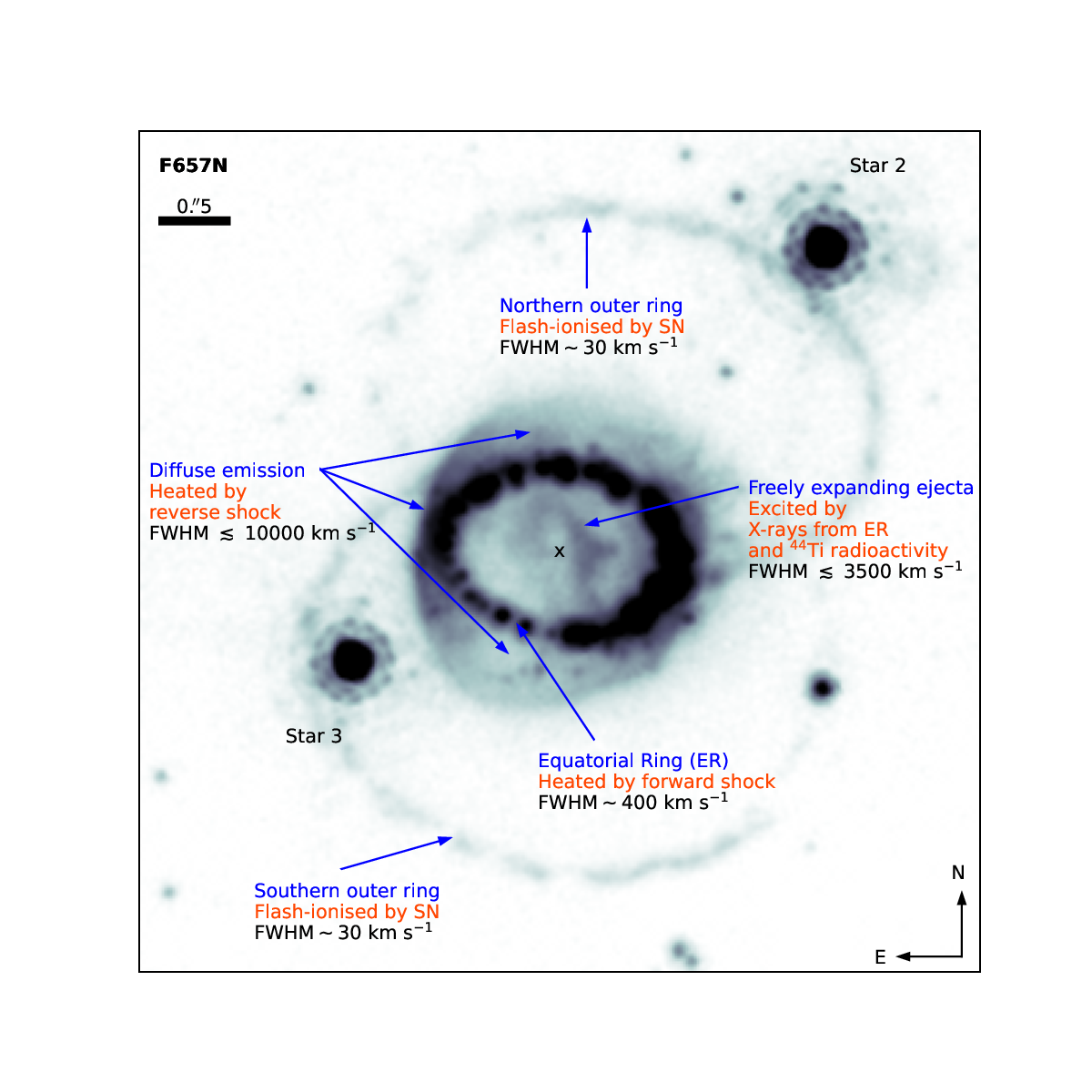}
\caption{HST/WFC3 2022 image of SN\,1987A in the F657N filter scaled by an asinh function with labels showing the main emission components (dark blue) and corresponding processes responsible for these emissions (orange). The FWHM of the optical emission lines are indicated in black \citep[from][]{groningsson08, tziamtzis11, fransson13, larsson19a}. The cross represents the geometric centre of the ER \citep{alp18}. We note that Star 1 is Sanduleak-69°\,202, the star that exploded into SN\,1987. The field of view is $6\fds0 \times 6\fds0$. Figure adapted with permission from \citet{rosu24}.}
\label{fig:general}
\end{figure}

Supernova (SN) 1987A in the Large Magellanic Cloud (LMC) located 49.59\,kpc away \citep{pietrzynski19} is the closest naked-eye SN since Kepler's SN in 1604, making it  one of the most thoroughly studied astronomical object in the Universe. The SN event was first observed on 1987 February 23 \citep[see][for reviews]{kunkel87,arnett89,mccray93,mccray16}. Its evolution has notably been followed with the Hubble Space Telescope (HST) since 1990. The excellent spatial resolution offered by the HST images allowed us to resolve the expanding SN ejecta and their interaction with the circumstellar triple-ring nebula made of the inner equatorial ring (ER), and the fainter northern and southern outer rings (NOR and SOR, see Fig.\,\ref{fig:general}). We refer to \citet{rosu24} for the latest analysis of HST data focusing on the SN ejecta and ER.  

The ER and ORs have on sky radii of $\sim$$\,0\fds8$ and $\sim$$\,2\fds4$, respectively. The ER, NOR, and SOR have inclination angles of $\sim$$43^\circ$, $\sim$$45^\circ$, and $\sim$$38^\circ$, respectively \citep{tziamtzis11}. The ORs are offset from the equatorial plane \citep[their figure 5]{tziamtzis11}. The northern (resp. southern) parts of the rings are the nearest (resp. farthest) to us \citep{panagia91, plait95, sugerman05}. The formation of the rings is likely the outcome of a binary merger that happened $\sim$$20\,000$ years before the SN explosion \citep{morris07, morris09}, as established from their pre-shock expansion velocities \citep{crotts00}.

\begin{table}
\centering
\caption{Details of the HST observations.}
\begin{tabular}{llll}
\hline\hline
Date & Epoch$^a$ & Instrument$^b$/Filter & Exposure time \\
(yyyy mmm dd) & (days) & & (s) \\
\hline
1994 Feb 03 & ~~2537 & WFPC2/F502N & 2400 \\
1996 Feb 06 & ~~3270 & WFPC2/F502N & 7800 \\
1997 Jul 12 & ~~3792 & WFPC2/F502N & 8200 \\
2000 Jun 16 & ~~4862 & WFPC2/F502N & 3600 \\
2000 Nov 14 & ~~5013 & WFPC2/F502N & 5600 \\
2001 Dec 07 & ~~5401 & WFPC2/F502N & 4800 \\
2003 Jan 05 & ~~5795 & ACS/F502N & 4000 \\
2003 Nov 28 & ~~6122 & ACS/F502N & 4000 \\
2004 Dec 15 & ~~6505 & ACS/F502N &  3600 \\
2005 Nov 18 & ~~6843 & ACS/F502N &  3180 \\
2006 Dec 08 & ~~7228 & ACS/F502N &  2600 \\
2009 Apr 29 & ~~8101 & WFPC2/F502N & 6900 \\
2014 Jun 20 & ~~9979 & WFC3/F502N & 5880 \\
2016 Jun 08 & 10\,698 & WFC3/F502N & 2400 \\
2018 Jul 10 & 11\,460 & WFC3/F502N & 5760 \\
2020 Aug 09 & 12\,221 & WFC3/F502N & 5600 \\
2021 Aug 23 & 12\,600 & WFC3/F502N & 5600 \\
2022 Sep 07 & 12\,980 & WFC3/F502N & 5600 \\
2009 Dec 13 & ~~8329 & WFC3/F657N & 1600 \\
2011 Jan 05 & ~~8717 & WFC3/F657N & 2400 \\
2016 Jun 08 & 10\,698 & WFC3/F657N & 2400 \\
2017 Aug 03 & 11\,119 & WFC3/F657N & 2800 \\
2018 Jul 10 & 11\,460 & WFC3/F657N & 2880 \\
2019 Jul 24 & 11\,839 & WFC3/F657N & 2880 \\
2020 Aug 09 & 12\,221 & WFC3/F657N & 2880 \\
2021 Aug 21 & 12\,598 & WFC3/F657N & 2800 \\
2022 Sep 06 & 12\,979 & WFC3/F657N & 2600 \\
\hline
\end{tabular}
\begin{tablenotes}
\item\textbf{Notes.} $^a$Epoch is measured in number of days since the explosion on 1987 February 23. $^b$WFC3 images were taken with the UVIS2 chip except for the 2009 observation taken with the UVIS1 chip.
\end{tablenotes}
\label{table:HSTobs}
\end{table} 

While the SN ejecta and ER have been thoroughly studied during the last four decades, the latest analysis of the ORs dates back to the last decade and is attributed to \citet{tziamtzis11}. In the current paper, we revisit the ORs of SN\,1987A in light of the HST data gathered from 1994 to 2022 in narrow-band filters and the newly acquired VLT/Multi-Unit Spectroscopic Explorer (MUSE) Integral Field Unit (IFU) observations. We provide the first infrared (IR) imaging and spectra, from James Webb Space Telescope (JWST) Near InfraRed Camera (NIRCam) and Mid-InfraRed Instrument (MIRI) Medium Resolution Spectrograph (MRS) IFU observations. We provide three decades of light curves of the ORs from HST. We specifically used two narrow-band filters, centered on H$\alpha$ and [O\,\textsc{iii}] $\lambda$\,5007, motivated by the fact that the ORs are known to emit narrow lines. We present the first analysis of the evolution in morphology of the ORs in the optical, and the first complete mid-resolution spectrum of the ORs in the optical and mid-infrared (MIR).

The paper is organised as follows. We describe the observations and data reduction processes in Sect.\,\ref{sect:observations}. The change of morphology of the ORs and the light curves from days 2537 to 12\,980\footnote{Hereafter, `days' refer to days after the explosion on 1987 February 23.} after the explosion in the narrow-band HST filters F502N and F657N are presented in Sect.\,\ref{sect:HST}. The MUSE and JWST data are analysed in Sects\,\ref{sect:MUSE} and \ref{sect:JWST}, respectively. The results are discussed in Sect.\,\ref{sect:discussion}, and our conclusions provided in Sect.\,\ref{sect:conclusion}. Throughout this paper, we refer to spectral lines by their air wavelengths in the optical and vacuum wavelengths in the IR (to follow standard conventions).


\begin{figure*}
\includegraphics[width=\linewidth]{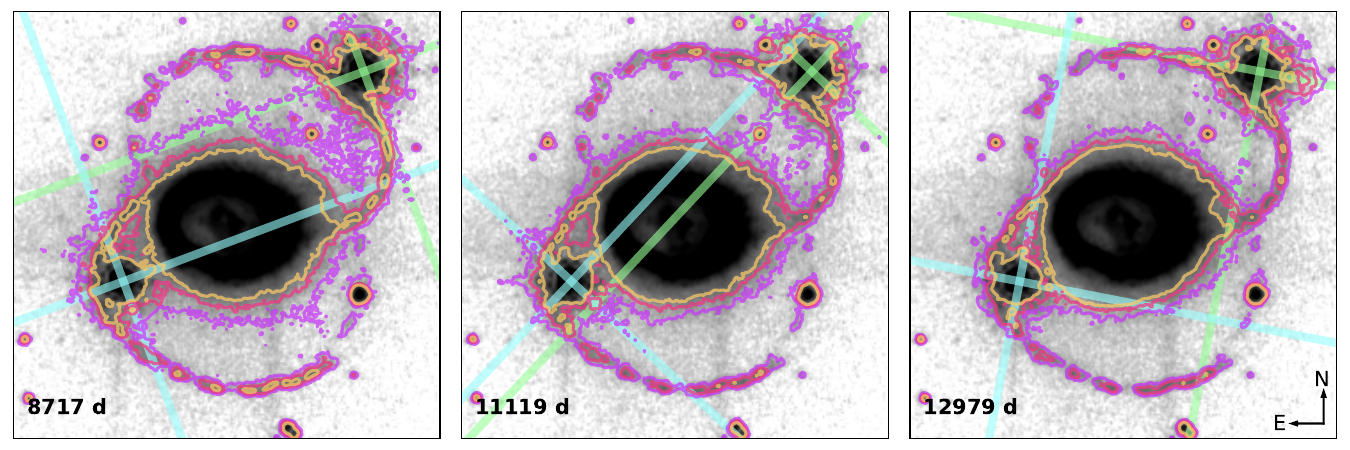}
\caption{HST/WFC3 contour images of SN\,1987A taken in the F657N filter at 8717, 11\,119, and 12\,979 days after the explosion over-plotted on the HST/WFC3 image taken at day 8329. Levels vivid orchid -- wild strawberry -- jasmine are of increasing flux (4\%, 8\%, and 14\%, respectively, of the maximum level), the same in all images, highlighting broken clumps. The light green and light blue lines represent the diffraction spikes from Stars 2 and 3, respectively. The field of view for each image is $6\fds0 \times 6\fds0$.}
\label{fig:contours_F657N}
\end{figure*}

\begin{table}
\centering
\caption{JWST/MIRI/MRS spatial and spectral sample dimensions.}
\begin{tabular}{llll}
\hline\hline
Channel/Band &Field of view & Pixel size & Wavelengths \\
 & (arcsec) & (arcsec) & ($\mu$m)  \\
\hline
1 Long (1L) &$3\fds2$$\times$$3\fds7$ & $0.196$ & $6.53-7.65$ \\
2 Short (2S) &$4\fds0$$\times$$4\fds8$ & $0.196$ & $7.51-8.77$ \\
2 Medium (2M) &$4\fds0$$\times$$4\fds8$ & $0.196$ & $8.67-10.13$ \\
2 Long (2L) &$4\fds0$$\times$$4\fds8$ & $0.196$ & $10.02-11.70$ \\
3 Short (3S) &$5\fds2$$\times$$6\fds2$ & $0.245$ & $11.55-13.47$ \\
3 Medium (3M) &$5\fds2$$\times$$6\fds2$ & $0.245$ & $13.34-15.57$ \\
3 Long (3L) &$5\fds2$$\times$$6\fds2$ & $0.245$ & $15.41-17.98$ \\
4 Short (4S) &$6\fds6$$\times$$7\fds7$ & $0.273$ & $17.70-20.95$ \\
4 Medium (4M) &$6\fds6$$\times$$7\fds7$ & $0.273$ & $20.69-24.48$ \\
4 Long (4L) &$6\fds6$$\times$$7\fds7$ & $0.273$ & $24.19-27.90$ \\
\hline
\end{tabular}
\begin{tablenotes}
\item \textbf{Notes.} From \citet{wells15, argyriou23}.
\end{tablenotes}
\label{table:JWST_channels}
\end{table} 

\section{Observations and data reduction\label{sect:observations}}
The HST Wide Field and Planetary Camera 2 (WFPC2), Advanced Camera for Surveys (ACS), and Wide Field Camera 3 (WFC3)  observations analysed in this paper are detailed in Table\,\ref{table:HSTobs}. The WFPC2 and ACS observations were discussed in \citet{tziamtzis11} to which we refer for further information. The dithered exposures were combined using \texttt{DrizzlePac} \citep{fruchter10} adopting a pixel scale of $0\fds025$. Cosmic rays were removed and distortion corrections were applied as part of the drizzling process \citep{hoffmann21}. The HST astrometry was corrected by aligning field stars with Gaia as in \citet{rosu24}. The images were aligned to the same pixel frame using \texttt{astropy reproject} with bilinear interpolation.

The MUSE \citep{bacon10} IFU observations were taken on 2023-03-26 (day 13182) in the narrow-field mode with adaptive optics (PID 110.25AR.001, PI: C. Fransson). The total exposure time was 7836 s. The spectra cover the wavelength range from 4750\,\AA~to 9350\,\AA, with a gap between 5760\,\AA~and 6010\,\AA. The spectral resolution is $R=2595$ at central wavelength ($R=1740$ at 4800\,\AA~and 3450 at 9300\,\AA). The field of view is $7\fds42 \times 7\fds43$ with a sampling of $0\fds025$ per spatial pixel (spaxel), and covers both ORs. The spectral sampling is 1.25\,\AA~ per pixel. The MUSE data were processed using the ESO pipeline \citep{weilbacher20} for bias subtraction, flat fielding, distortion corrections, wavelength calibration, and cube combination. The observations used a dither pattern to improve sampling and remove instrument systematics. As the location of the supernova in the LMC does not provide nearby pointings with clean sky and the supernova region itself contains significant emission lines, the removal of the ‘sky’ background was performed using an iterative technique. From the reduction stage prior to the final sky subtraction we ‘manually’ constructed a ‘background’ spectrum which included sky lines, LMC emission, and scattered light. This spectrum was subsequently used by the MUSE pipeline to remove the foreground and leave us with a ‘clean’ SN\,1987A cube.

JWST NIRCam deep images of SN\,1987 were obtained on 2022-09-01 and 2022-09-02 (days 12974 and 12975) in four filters in the short wavelength channel (F150W, F164N, F200W, and F212N) and four filters in the long wavelength channel (F323N, F356W, F405N, F444W) under PID 1726 (PI: M. Matsuura). The field of views are larger than HST field of view and cover both ORs. We refer to \citet{matsuura24} for the observing strategy, reduction, and image convolution of the data.

\begin{figure*}
\includegraphics[clip=true, trim=0 25 80 25,width=\linewidth]{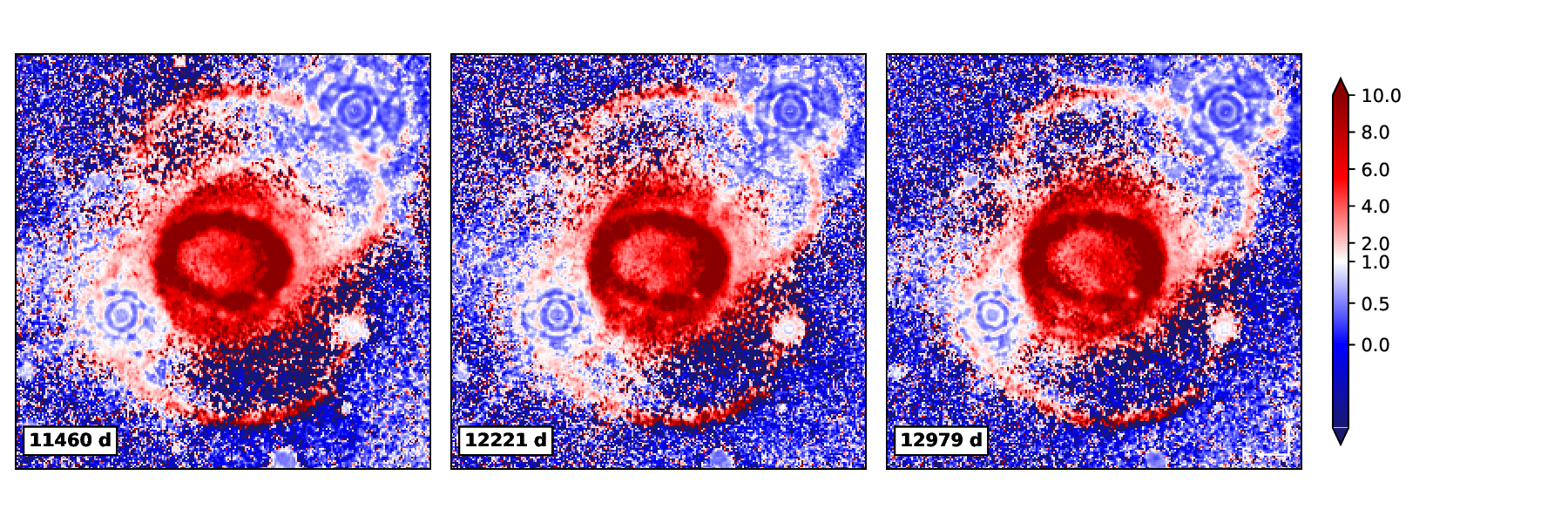}
\caption{HST/WFC3 F657N/F502N ratio images of SN\,1987A at 11\,460, 12\,221, and 12\,979 days after the explosion. The field of view for each image is $6\fds0 \times 6\fds0$.}
\label{fig:contours_F502N_F657N}
\end{figure*}

JWST MIRI/MRS IFU observations were obtained in Cycle 1 (PID 1232, PI: G. Wright) on 2022-07-16 (day 12927) and in Cycle 2 (PID 2763, PI: M. Meixner) on 2024-08-04 (day 13677). Observations consist of 94 groups of three integrations using \texttt{FASTR1} readout pattern for all three bands (short, medium, and long). The same configurations were adopted for both cycles. The total exposure time amounts to 3152.4 s per MIRI/MRS channel/band combination. The spatial and spectral sample dimensions are given in Table\,\ref{table:JWST_channels} \citep{wells15, argyriou23}.
We refer to figure 1 in \citet{kavanagh25} that shows the locations of the field of views on SN\,1987A. We refer to \citet{jones23b} for the description and analysis of Cycle 1 data and to \citet{kavanagh25} for the description, reduction, and analysis of Cycle 2 data as well as the re-reduction of Cycle 1 data. We here combined both cycles together using the \citet{kavanagh25} reduction as this gives both consistency and the best calibrations, and increases the S/N in the faint outer rings. There is no obvious time variation between Cycles 1 and 2 for the ER \citep{kavanagh25} so we assumed that it is the case for the ORs, too.


\section{Hubble Space Telescope observations\label{sect:HST}}
The HST observations analysed in this paper are shown in Appendix\,\ref{appendix:HST_images} (see Figs \ref{fig:obs_F502N1}, \ref{fig:obs_F502N2}, and \ref{fig:obs_F657N}).

\subsection{Morphology\label{subsect:morphology_HST}}
The ORs are only faintly distinguishable in the WFPC2/F502N observations taken at days 4862, 5013, 5401, and 8101, certainly due to the lower efficiency of the camera at later epochs. The ORs are brighter in the F657N filter than in the F502N filter because the H$\alpha$ line contributing to the F657N filter (that also covers [N\,\textsc{ii}] lines) is much stronger than the [O\,\textsc{III}] $\lambda$\,5007 line contributing to the F502N filter. From the images themselves, no drastic change in the morphology of the ORs with time stands out, in either of the two filters. 

\begin{figure}
\centering
\includegraphics[width=\linewidth]{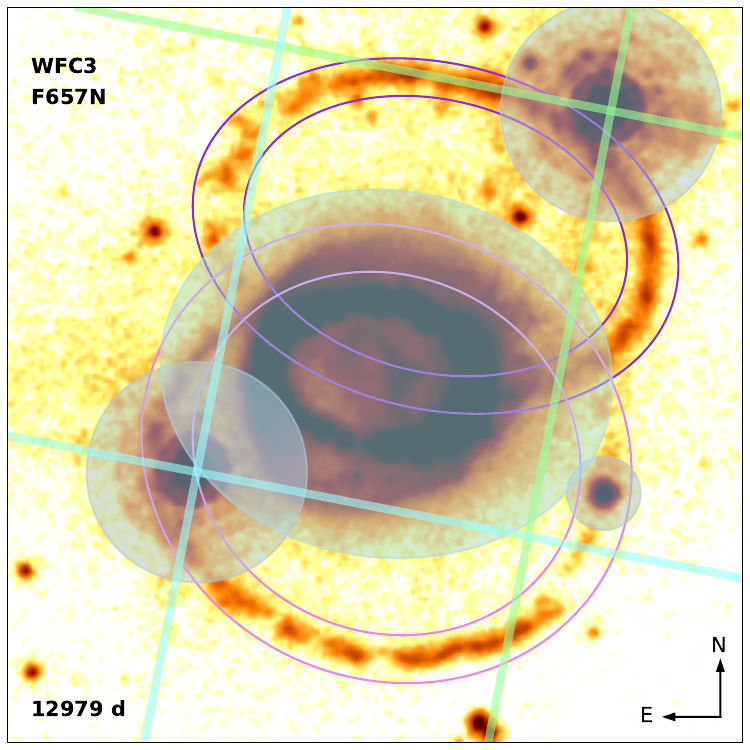}
\caption{HST/WFC3 2022 image of SN 1987A in the F657N filter together with the regions adopted to compute the fluxes. The dark and light purple lines define the elliptical annuli around the NOR and SOR, respectively. The light blue shaded area define the regions contaminated by the SN ejecta, the ER, the reverse shocks, Star 2, Star 3, and the star located in the western part of the SOR. Diffraction spikes colours are as in Fig.\,\ref{fig:contours_F657N}. The field of view is $6\fds0 \times 6\fds0$.  \label{fig:regions}}
\end{figure}

In order to better highlight small changes in morphology with time in the F657N filter, we present in Fig.\,\ref{fig:contours_F657N} contours plots of three HST images taken at days 8717, 11119, and 12979 over-plotted on the HST image taken at day 8329. Both the NOR and SOR have become less bright with time, as each clump in the ORs has become dimmer (see Sect.\,\ref{subsect:lightcurves} for a quantitative analysis). This is especially evident for the SOR. Some parts of the ORs close to Stars 2 and 3 seem to have become brighter at some observations: this is an artefact from the diffraction spikes of the stars (see green and blue lines in Fig.\,\ref{fig:contours_F657N}). The same general behaviour is observed in the F502N filter (not shown).

In Fig.\,\ref{fig:contours_F502N_F657N}, we show the ratio between HST/WFC3 images taken in the F657N and F502N filters at the same epoch for three epochs. Qualitatively, the brightness of the ORs in the F657N filter has decreased faster than in the F502N filter: the ORs appear slightly redder in the first ratio image than in the last (see also Sect.\,\ref{subsect:lightcurves}).

\subsection{Light Curves\label{subsect:lightcurves}}
We analysed the light curves of the ORs in the two HST filters F502N and F657N from day 2537 to day 12\,980. The fluxes in the NOR and SOR were computed assuming the same elliptical annuli for all observations, that is to say, assuming the expansion velocity of the ORs is negligible\footnote{The expansion velocity of the ORs measured with respect to the SN was determined by \citet{crotts00}: $\sim$$20$$-$$25$\,km\,s$^{-1}$, corresponding to 0.1 mas\,yr$^{-1}$, i.e. 0.004 HST/WFC3 pixels per year. Over the 28.6 yr from day 2537 to 12\,980, this would total to maximum 0.13 pixels of expansion.}. The regions were chosen to be sufficiently large to encompass the bright emission seen in all observations, and are represented by the purple lines in Fig.\,\ref{fig:regions}. The NOR (resp. SOR) annulus has an aspect ratio of 0.7 (resp. 0.9), a position angle of $166^\circ$ (resp. $154^\circ$) measured by the position of the major axis in the trigonometric circle, and its internal and external major axes equal to 127 and 161 (resp. 129 and 163) pixels, corresponding to $3\fds2$ and $4\fds0$ (resp. $3\fds2$ and $4\fds1$), respectively. We discarded the regions contaminated by either the SN ejecta, the ER, the reverse shocks, Star 2, Star 3, or the star located in the western part of the SOR (Fig.\,\ref{fig:general}). We adopted conservative exclusion regions: they are encircled in light blue in Fig.\,\ref{fig:regions}.

\begin{figure}
\centering
\includegraphics[width=\linewidth]{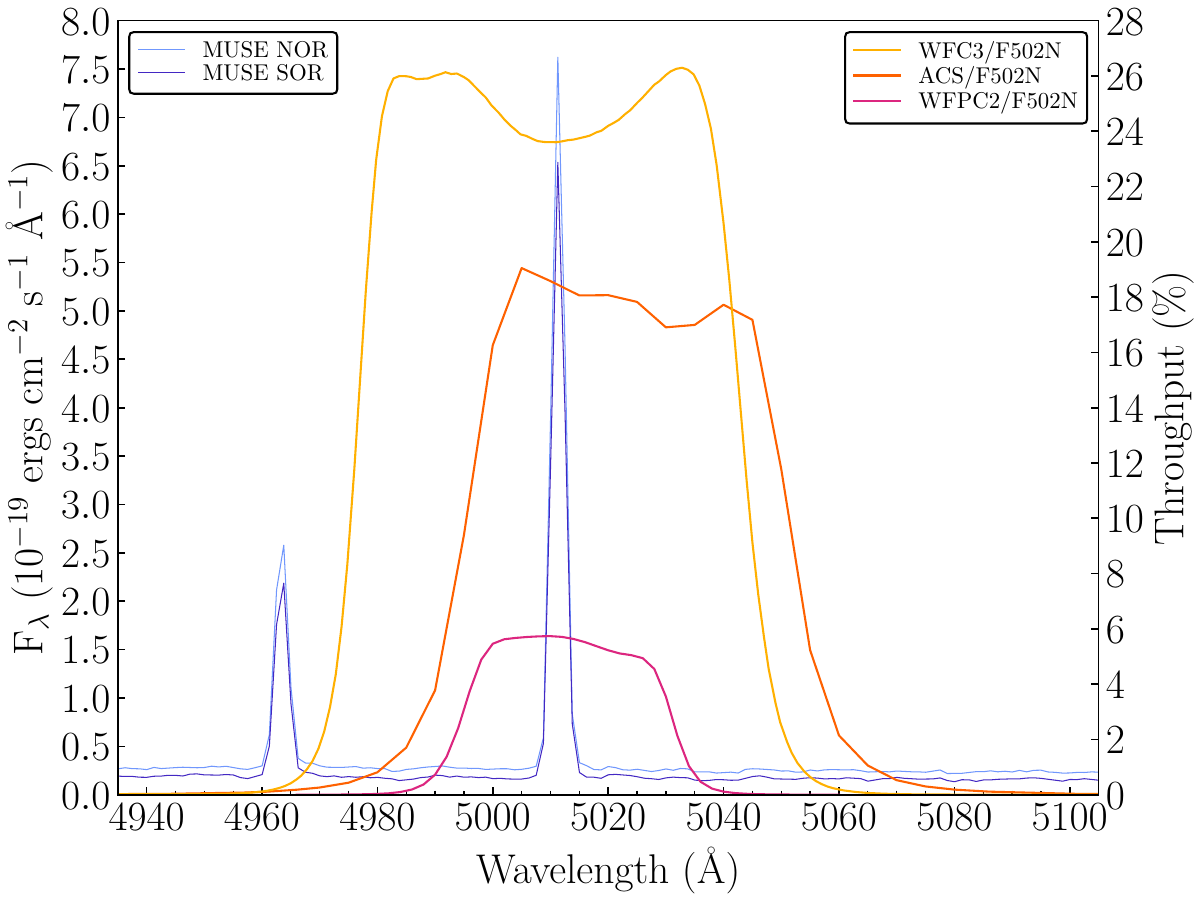}
\includegraphics[width=\linewidth]{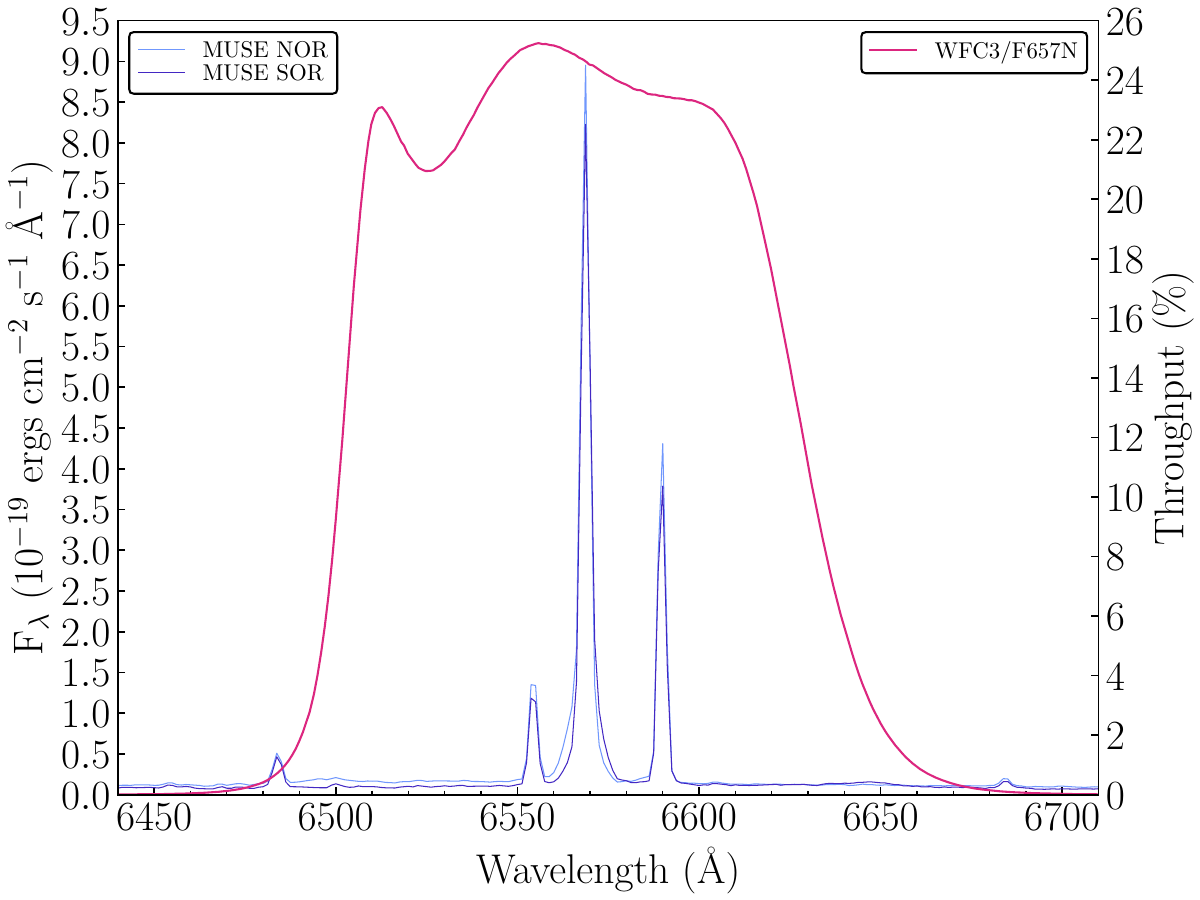}
\caption{MUSE not background subtracted spectra of the NOR and SOR described in Sect.\,\ref{sect:MUSE} and HST/WFPC2, ACS, and WFC3 F502N filters’ response functions (top panel), and WFC3 F657N filter's response function (bottom panel). The F502N filters cover the [O\,\textsc{iii}] $\lambda$\,5007 line, only the WFC3/F502N marginally covers the [O\,\textsc{iii}] $\lambda$\,4959 line. The F657N covers the H$\alpha$ line and the two neighbouring [N\,\textsc{ii}] $\lambda\lambda$\,6548, 6563 lines. The line at $\sim$\,6480\,\AA~is from the background.}
\label{fig:convolution}
\end{figure}

Given that the HST observations were performed with three different instruments (WFPC2, ACS, and WFC3) for which the relevant filter F502N differs in terms of both wavelength coverage and throughput, a direct comparison of fluxes is difficult. To account for the differences between the three different instruments used, we computed multiplicative correction factors to be applied to the WFPC2 and ACS fluxes to be compared to the WFC3 fluxes. We determined these corrections factors by convolving the MUSE spectra of the NOR and SOR (presented in Sect.\,\ref{sect:MUSE}) with the response functions of the different instrument/filter combinations (see Fig.\,\ref{fig:convolution}). We adopted as correction factors the ratio between the obtained convoluted fluxes in the WFPC2 or ACS and the WFC3. The correction factors for WFPC2/WFC3 are of 6.20 and 5.99 for the NOR and SOR, respectively. The correction factors for ACS/WFC3 are of 1.42 and 1.39 for the NOR and SOR, respectively. The response functions of the WFC3/F657N UVIS1 and UVIS2 chips are equal. All light curves presented in this paper were corrected by these factors and hence normalised to the bandpass of the WFC3 filter. The largest corrections apply to the WFPC2 F502N filter which is significantly narrower than the corresponding filters used with the WFC3 and ACS. The correction factors are slightly larger for the NOR than the SOR because the red wing of the [O\,\textsc{III}] $\lambda$\,4959 line is slightly contributing in the NOR in the WFC3 but is not contributing in the WFPC2 and ACS, nor in the SOR, and because the continuum is higher in the NOR than in the SOR. Uncertainty exists in these correction factors due to the time evolution of the ORs spectra, though given their slow time evolution, we estimate  these uncertainties to be minimal. Indeed, \citet{larsson19a} showed that the spectral evolution of the ER and SN ejecta affects the correction factors by less than 1\% over the same timespan, hence the slower spectral evolution of the ORs should have a significant smaller impact.

\setlength{\tabcolsep}{1.5mm}
\begin{table}
\caption{De-reddened flux measurements in the ORs in all HST epochs and filters. \label{table:fluxes_HST}}
\centering
\begin{tabular}{llll}
\hline\hline
Epoch & Filter & NOR Flux$^a$ & SOR Flux$^{a,b}$ \\
(days) & & ($10^{-14}$ erg\,cm$^{-2}$\,s$^{-1}$) & ($10^{-14}$ erg\,cm$^{-2}$\,s$^{-1}$) \\
\hline
~~2537 & F502N & $8.017\pm 0.207$ & $5.558\pm 0.170$ \\
~~3270 & F502N & $3.602\pm 0.077$ & $1.885\pm 0.055$ \\ 
~~3792 & F502N & $2.590\pm 0.064$ & $0.765\pm 0.034$ \\
~~5795 & F502N & $0.771\pm 0.005$ & $0.229\pm 0.003$ \\
~~6122 & F502N & $0.634\pm 0.005$ & $0.126\pm 0.002$ \\
~~6505 & F502N & $0.623\pm 0.005$ & $0.197\pm 0.003$ \\
~~6843 & F502N & $0.725\pm 0.005$ & $0.118\pm 0.002$ \\
~~7228 & F502N & $0.681\pm 0.006$ & $0.165\pm 0.003$ \\
~~9979 & F502N & $0.528\pm 0.002$ & $0.115\pm 0.001$ \\
10\,698 & F502N & $0.497\pm 0.004$ & $0.112\pm 0.002$ \\
11\,460 & F502N & $0.530\pm 0.003$ & $0.120\pm 0.001$ \\
12\,221 & F502N & $0.490\pm 0.002$ & $0.132\pm 0.001$ \\
12\,600 & F502N & $0.485\pm 0.002$ & $0.064\pm 0.001$ \\
12\,980 & F502N & $0.464\pm 0.002$ & $0.073\pm 0.001$ \\
~~8329 & F657N & $1.767\pm 0.007$ & $1.454\pm 0.006$ \\
~~8717 & F657N & $1.870\pm 0.006$ & $1.437\pm 0.005$ \\
10\,698 & F657N & $1.609\pm 0.005$ & $1.165\pm 0.004$ \\
11\,119 & F657N & $1.602\pm 0.005$ & $1.067\pm 0.004$ \\
11\,460 & F657N & $1.581\pm 0.005$ & $1.093\pm 0.004$ \\
11\,839 & F657N & $1.524\pm 0.005$ & $1.090\pm 0.004$ \\
12\,221 & F657N & $1.441\pm 0.005$ & $1.072\pm 0.004$ \\
12\,598 & F657N & $1.367\pm 0.004$ & $0.973\pm 0.004$ \\
12\,979 & F657N & $1.329\pm 0.005$ & $0.934\pm 0.004$ \\
\hline
\end{tabular}
\begin{tablenotes}
\item\textbf{Notes.} $^a$Fluxes were summed over the pixels encompassed in the regions defined in Sect.\,\ref{subsect:lightcurves}. Fluxes in the WFPC2 and ACS cameras were corrected as explained in Sect.\,\ref{subsect:lightcurves}. $^b$The SOR fluxes were scaled to the NOR regions as explained in Sect.\,\ref{subsect:lightcurves}. 
\end{tablenotes}
\end{table}

\begin{figure}
\includegraphics[width=\linewidth]{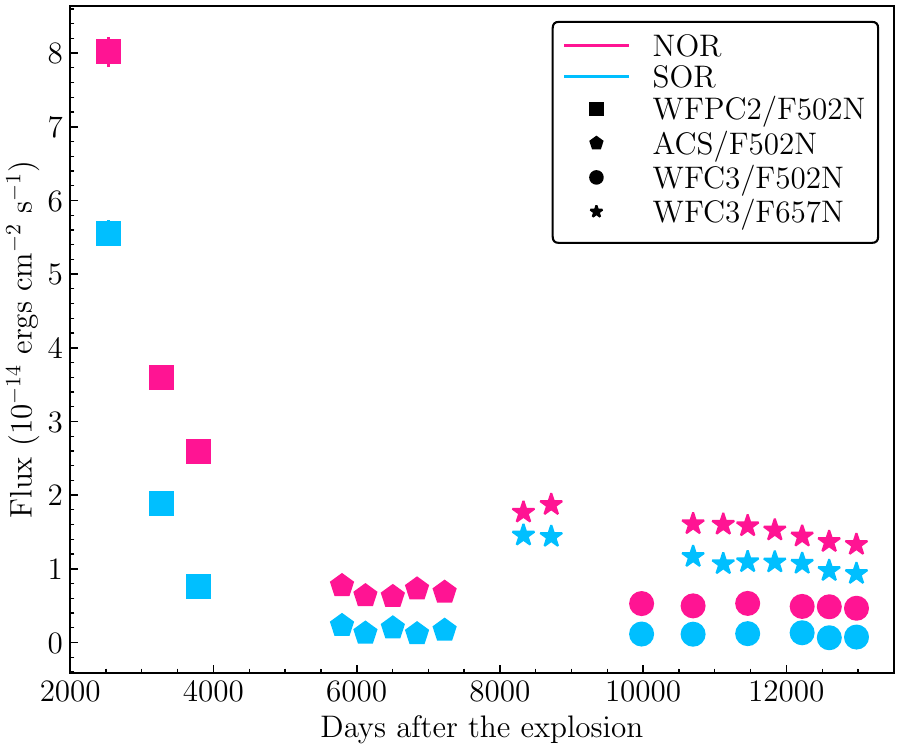}
\caption{Light curves of the NOR and SOR in the HST F502N and F657N filters. The WFPC2/F502N and ACS/F502N fluxes were corrected as explained in Sect.\,\ref{subsect:lightcurves}. The $1\sigma$ statistical uncertainties are smaller than the symbols. \label{fig:lightcurve}}
\end{figure}

As pointed out by \citet{larsson19a}, the WFPC2/F502N observation at day 8101 is significantly affected by the degradation of the charge transfer efficiency (CTE), meaning that lower fluxes were measured due to the trapping of charges during CCD readout. In addition, we measured negative fluxes in the SOR for the WFPC2/F502N observations at days 4862, 5013, and 5401 (the ORs do not stand out of the background in these observations, see Sect.\,\ref{subsect:morphology_HST} and Fig.\,\ref{fig:obs_F502N1}). As explained in \citet{larsson19a}, while correction formulas for CTE losses exist for isolated point sources, these are not appropriate for SN\,1987A, which consists of a combination of diffuse emission and closely spaced point sources. Therefore, we decided to discard these observations for the light curve analysis. 

We summed the count rates over the pixels inside the given regions (a pixel is considered inside the region if its centre is strictly inside that region). The $1\sigma$ statistical and systematic uncertainties are negligible. The count rates were converted into fluxes using the inverse sensitivity of the filters. Finally, the fluxes were de-reddened adopting the extinction curve of \citet{maiz14} with a colour excess $E(B-V)= 0.19$\,mag and a reddening factor in the V-filter $R_V = 3.1$, as suggested by \citet{france11}.

Depending on the orientation of the telescope at the time of the observation, diffraction spikes from Stars 2 and 3 cross the NOR and SOR regions differently (see example in Fig.\,\ref{fig:regions}). We accounted for the extra contribution of the diffraction spikes from Stars 2 and 3 to the flux in the NOR and SOR regions by subtracting it as described in \citet[their Appendix D]{rosu24}. This correction is at the level of 17\% at most, except for one deviant point. The reduction of the fluxes due to the diffraction spikes is given in Appendix\,\ref{appendix:diffraction_spikes}. The negative flux reductions are within the noise level.

\begin{figure*}[t!]
\includegraphics[width=0.97\linewidth]{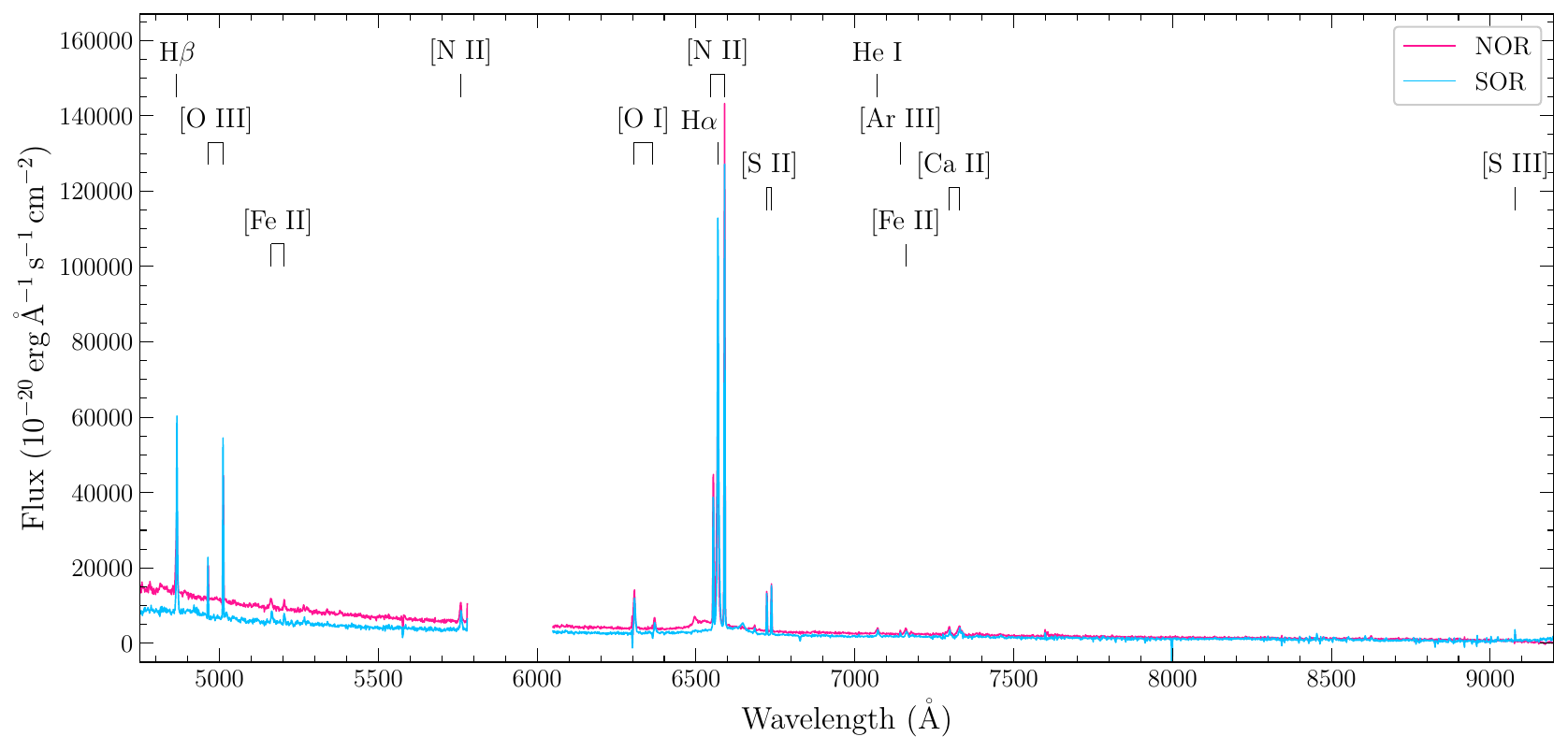}
\caption{MUSE mean background subtracted spectra of the NOR and SOR with main lines identified. The line at $\sim$\,$6500$\,\AA~in the NOR is contamination from the reverse shocks.
\label{fig:spectra_MUSE}}
\end{figure*}

Given that the number of pixels in the NOR and SOR are different (ratio of 1.02 in surface area covered), we scaled the SOR by the ratio between the number of pixels in the NOR and SOR to allow for a quantitative comparison between both ORs. The flux measurements in the NOR and SOR in all HST observations are provided in Table\,\ref{table:fluxes_HST}. Both statistical and systematic errors are accounted for; For the latter, we propagated the uncertainties from the spatial variations of the background. The light curves are presented in Fig.\,\ref{fig:lightcurve}. The flux is systematically lower in the SOR than in the NOR in both filters. The flux decreases in both filters with time. The rate of decrease in flux was especially fast between 2537 and 3792 days after the explosion, but slowed down afterwards. This is due to the UV flash from the SN that excited the ORs right after the explosion, and the ORs just fading since then. The bumps of brightening observed after 6000\,days are unlikely to be real.
At the latest epochs, the rate of decrease in flux is marginally steeper in the F657N filter than in the F502N, as already observed in Sect.\,\ref{subsect:morphology_HST}, Fig.\,\ref{fig:contours_F502N_F657N}. The flux in F502N at the latest epochs seem to flatten. From this evolution, we rule out the possibility that the SN ejecta have already interacted with the ORs. 

We further sampled the NOR and SOR regions each into 10 small regions to investigate whether there is any difference in the evolution depending on the location on the ORs. We observed the same general trend as the full corresponding NOR and SOR regions, but did not reveal any difference between sub-regions: the small variations we observed between two consecutive epochs were attributed to either the low total flux of the regions and lower statistics or the proportionally more significant contribution of the diffraction spikes. We only confirmed what we see in the images: the flux in the western part of the NOR is larger than in the eastern part of the NOR, but both decrease at the same rate with time. The same is observed for the eastern part of the SOR compared to its western part. The wide HST filters are not as suited for this purpose as the MUSE data are (see Sect.\,\ref{sect:MUSE}). 


\section{MUSE observations\label{sect:MUSE}}
We extracted MUSE spectra for the NOR and SOR adopting the same regions as for the HST analysis (see Fig.\,\ref{fig:regions}). The mean spectra and main identification lines are shown in Fig.\,\ref{fig:spectra_MUSE}. All lines are symmetric except for H$\alpha$: it has a stronger blue- (resp. red-) shifted contribution in the NOR (resp. SOR) than in the SOR (resp. NOR) which is due to contamination from the reverse shock, blue-shifted in the north and red-shifted in the south \citep{larsson23}. 

\begin{figure}[t]
\includegraphics[width=\linewidth]{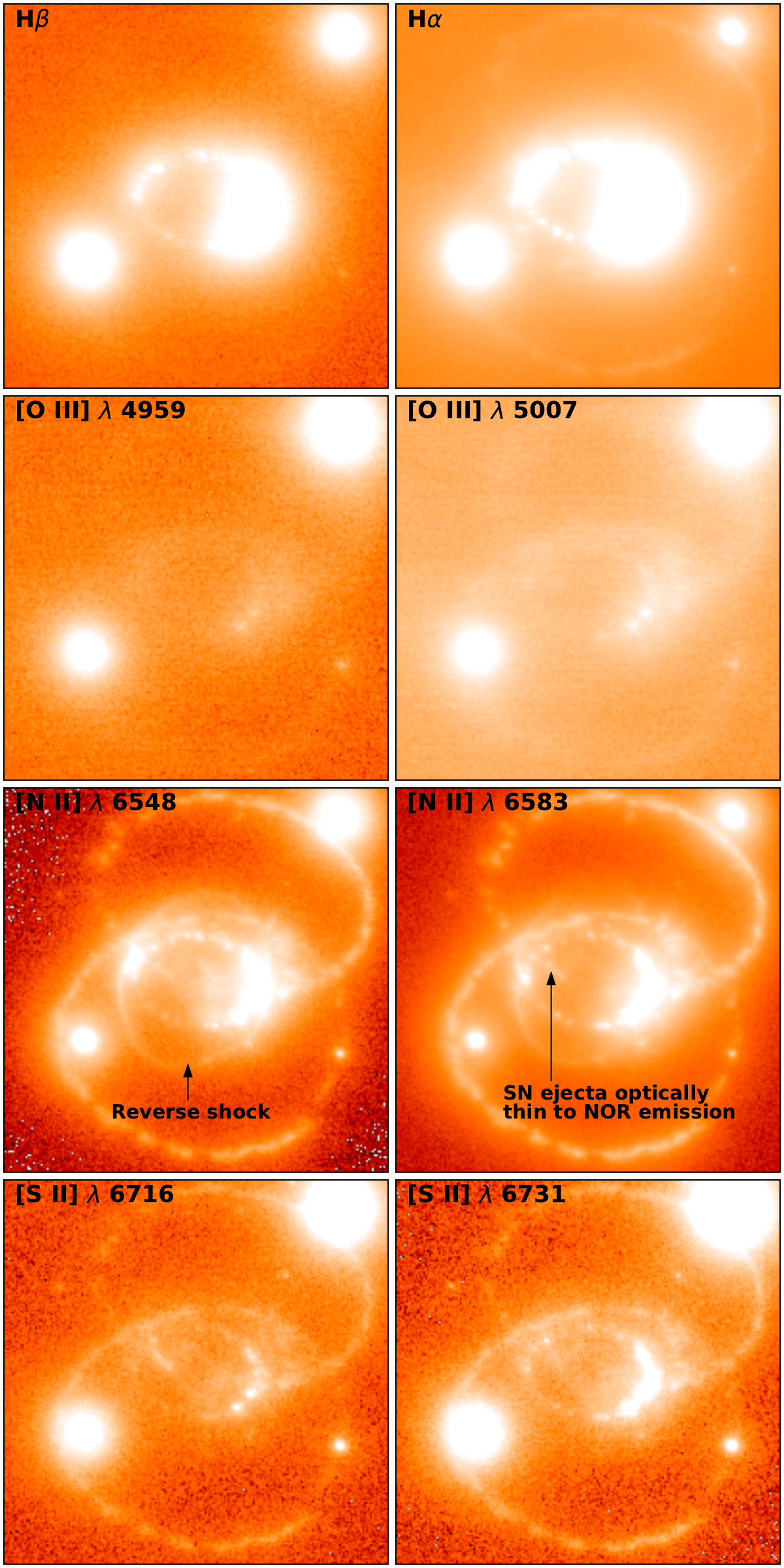}
\caption{MUSE slab images around the main emission components of the ORs (see Fig.\,\ref{fig:spectra_MUSE}). Only those where we can clearly see the ORs are shown. The additional thin rings observed in [N\,\textsc{ii}] images are H$\alpha$ contributions from the reverse shocks. The field of view for each image is $5\fds0 \times 5\fds0$.\label{fig:images_MUSE}}
\end{figure}

\begin{figure}
\centering
\includegraphics[clip=true,trim=0 20 0 20,width=\linewidth]{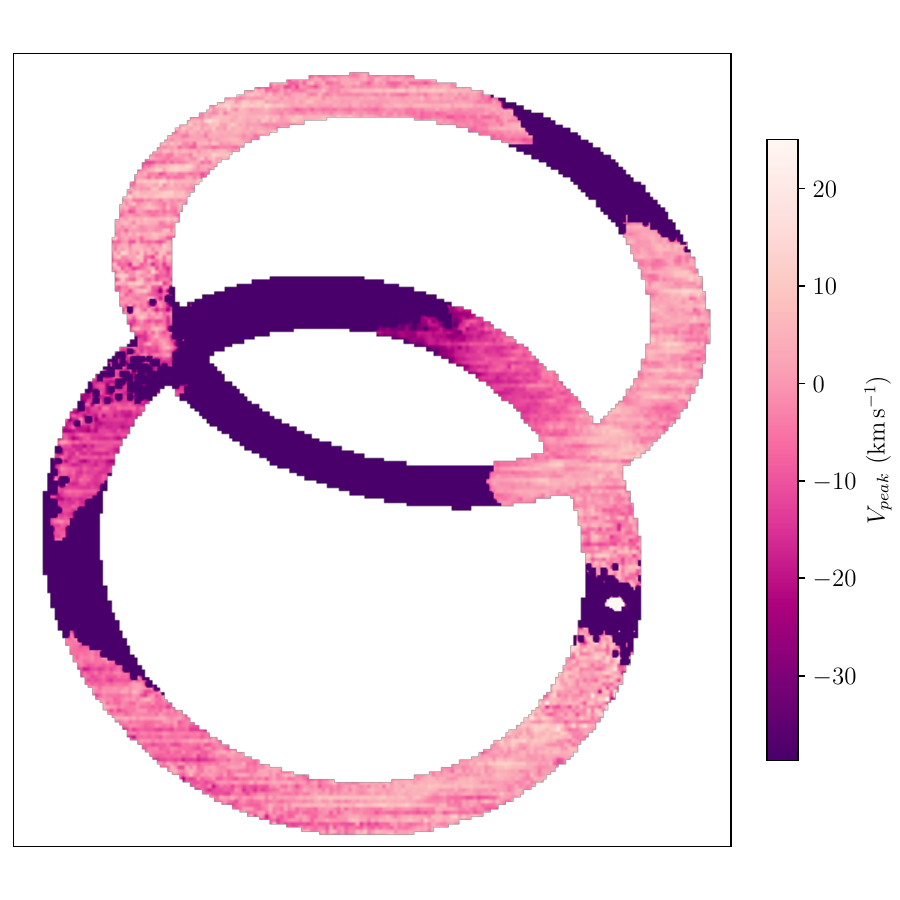}
\caption{Colourmaps in the ORs of the velocity at maximum flux ($V_\text{peak}$) of the [N\,\textsc{ii}] $\lambda$\,6583 line. The field of view is $5\fds0 \times 5\fds0$.\label{fig:MUSE_lambda_and_fwhm}}
\end{figure}

In Fig.\,\ref{fig:images_MUSE} we show MUSE slab images around the main emission components. The ORs are best seen in the [N\,\textsc{ii}] $\lambda\lambda$\,6548, 6583, and [S\,\textsc{ii}] $\lambda\lambda$\,6716, 6731 lines -- while these lines are not the strongest in the ORs -- because the spatial resolution is best at these wavelengths due to the adaptive optics. The morphology of the ORs is the same in all filters. We do not observe any clump nor region brighter or dimmer than another one in another emission line. A striking feature is the appearance of the NOR at the location of the SN ejecta a bit east to the geometric centre of the ER (indicated in Fig.\,\ref{fig:images_MUSE}): as the southern part of the NOR is located behind the SN ejecta in our field of view, it means the SN ejecta are optically thin to this wavelength.

We computed the line-integrated flux of the emission lines in the NOR and SOR by summing over the pixels encompassed by the defined regions. The fluxes were de-reddened adopting the same extinction curve as for the HST observations (Sect.\,\ref{subsect:lightcurves}). As for the HST observations, we scaled the SOR by the ratio between the number of pixels in the NOR and SOR to allow for a quantitative comparison between both ORs. The flux measurements in the NOR and SOR in the MUSE emission lines are provided in Table\,\ref{table:MUSE}. We accounted for both statistical and systematic errors as for HST images (see Sect.\,\ref{subsect:lightcurves}).

\setlength{\tabcolsep}{1mm}
\begin{table}[h!]
\caption{Emission lines from the NOR and SOR as observed with MUSE in March 2023 (day 13182).   \label{table:MUSE}}
\centering
\begin{tabular}{llll}
\hline\hline
Emission & Rest $\lambda^a$ & NOR Flux$^b$ & SOR Flux$^{b,c}$  \\
& (\AA) & ($10^{-16}$\,erg\,cm$^{-2}$\,s$^{-1}$)  & ($10^{-16}$\,erg\,cm$^{-2}$\,s$^{-1}$) \\
\hline
H$\beta$ & 4861.32 & $12.539\pm0.017$ & $20.012\pm0.020$ \\
$[$O\,\textsc{iii}$]$ & 4958.91 & $2.616\pm0.003$ & $4.178\pm0.006$ \\
$[$O\,\textsc{iii}$]$ & 5006.84 & $7.439\pm0.006$  & $12.860\pm0.009$ \\
$[$Fe\,\textsc{ii}$]$& 5158.00 & $1.285\pm0.006$ & $1.418\pm0.005$ \\
$[$Fe\,\textsc{ii}$]$&5199.17 & $0.705\pm0.004$ & $0.886\pm 0.004$ \\
$[$N\,\textsc{ii}$]^d$ & 5754.59 & $2.617\pm0.003$ & $2.267\pm0.003$ \\
$[$O\,\textsc{i}$]^e$ & 6300.30 & ...  & ... \\
$[$O\,\textsc{i}$]^e$ & 6363.78 & ...  & ... \\
$[$N\,\textsc{ii}$]$ & 6548.05 & $13.576\pm0.050$ & $11.934\pm0.022$ \\
H$\alpha$ & 6562.80 & $39.634\pm0.062$ & $47.608\pm0.082$ \\
$[$N\,\textsc{ii}$]$ & 6583.45 & $38.673\pm0.017$  & $35.499\pm0.025$ \\
$[$S\,\textsc{ii}$]$ & 6716.44 & $3.074\pm0.002$  & $3.105\pm0.001$ \\
$[$S\,\textsc{ii}$]$ & 6730.82 & $3.402\pm0.001$ & $3.601\pm0.001$ \\
He\,\textsc{i} & 7065.71 & $1.138\pm0.002$ & $0.979\pm0.002$ \\
$[$Ar\,\textsc{iii}$]$ & 7135.79 & $0.224\pm0.001$ & $0.296\pm0.001$ \\
$[$Fe\,\textsc{ii}$]$& 7155.54& $1.140\pm0.002$ & $1.007\pm0.001$ \\
$[$Ca\,\textsc{ii}$]$& 7291.47& $1.628\pm0.003$ & $0.984\pm0.003$ \\
$[$Ca\,\textsc{ii}$]$&  7323.89& $2.623\pm0.004$ & $2.402\pm0.005$ \\
$[$S\,\textsc{iii}$]$& 9068.60& $0.463\pm0.002$ & $0.629\pm0.002$ \\
\hline
\end{tabular}
\begin{tablenotes}
\item\textbf{Notes.} $^a$Air wavelengths from the compilation by \citet{vanhoof18}. $^b$Fluxes were summed over the pixels encompassed in the regions defined in Sect.\,\ref{sect:MUSE}. $^c$The SOR fluxes were scaled to the NOR regions as explained in Sect.\,\ref{sect:MUSE}. $^d$This line is heavily contaminated by the surrounding medium; the values should be taken with caution. $^e$These lines are contaminated by artefacts and their flux cannot be measured.
\end{tablenotes}
\end{table}

\setlength{\tabcolsep}{1.3mm}
\begin{table}
\caption{De-reddened flux measurements in the ORs in all JWST/NIRCam filters. The strong lines falling into the NIRCam filter bands are given in the last column. \label{table:NIRCam_fluxes}}
\centering
\begin{tabular}{lllll}
\hline\hline
Filter & NOR Flux$^{a}$ & SOR Flux$^{a,b}$ & Lines\\
& \multicolumn{2}{c}{(10$^{-16}$ erg\,cm$^{-2}$\,s$^{-1}$)}   & ($\mu$m in vacuum)\\
\hline
F150W & $16.461 \pm 15.075 $ &  $9.916 \pm 16.164$ & [Fe\,\textsc{ii}] 1.257, 1.644\\
F164N & $1.994 \pm 0.745$ & $1.468 \pm 0.774$ & [Fe\,\textsc{ii}] 1.644\\
F200W & $16.039 \pm 11.021$ & $9.439 \pm 11.846$ & Br$\delta$, He\,\textsc{i} 2.059, Br$\gamma$\\
F212N & $0.456 \pm 0.520$ & $0.348 \pm 0.535$ & Field stars\\
F323N & $0.373 \pm 0.146$ & $0.305 \pm 0.215$ & Field stars\\
F356W & $5.817 \pm 1.102$ & $4.493 \pm 1.315$ & Continuum\\
F405N & $1.176 \pm 0.153$ & $1.121 \pm 0.298$ & Br$\alpha$\\
F444W & $6.537 \pm 0.973$ & $5.866 \pm 1.315$ & Br$\alpha$, Pf$\beta$\\
\hline
\end{tabular}
\begin{tablenotes}
\item\textbf{Notes.} $^a$Fluxes were summed over the pixels encompassed in the regions defined in Sect.\,\ref{subsect:nircam}. $^b$The SOR fluxes were scaled to the NOR regions as explained in Sect.\,\ref{subsect:nircam}. 
\end{tablenotes} 
\end{table}

\setlength{\tabcolsep}{0.9mm}
\begin{table}
\caption{Emission lines from the NOR and SOR as observed with JWST/MIRI/MRS.  \label{table:MRS_fluxes}}
\centering
\begin{tabular}{llll}
\hline\hline
Emission & Rest $\lambda^a$ & NOR Flux$^{b}$ & SOR Flux$^{b,c}$  \\
& ($\mu$m) & ($10^{-16}$ erg\,cm$^{-2}$\,s$^{-1}$)  & ($10^{-16}$ erg\,cm$^{-2}$\,s$^{-1}$) \\
\hline
$[$Ar\,\textsc{ii}$]$ & 6.9853 & $0.408 \pm 0.022$ & $0.494\pm0.024$\\
$[$Ne\,\textsc{vi}$]$ & 7.6524 & $0.511 \pm 0.020$&  $0.723\pm0.024$\\
$[$Ar \textsc{iii}$]$ & 8.9914 & $0.098 \pm 0.013$ & $0.035\pm0.012$ \\
$[$S \textsc{iv}$]$ & 10.5105 & $1.069 \pm 0.022$ & $0.397\pm0.022$ \\
$[$Ne \textsc{ii}$]$ & 12.8135 & $3.773 \pm 0.030$ & $2.584\pm0.043$ \\
$[$Ne \textsc{v}$]$ & 14.3217 & $0.986 \pm 0.020$ & $1.366\pm0.016$ \\
$[$Ne \textsc{iii}$]$ & 15.5551 & $6.956 \pm 0.044$ & $2.012\pm0.037$ \\
$[$S \textsc{iii}$]$ & 18.7130 & $3.167 \pm 0.022$ & $1.070\pm0.014$ \\
$[$Ne \textsc{v}$]$ & 24.3175 & $0.629\pm 0.044$ & $2.279\pm 0.057$ \\
$[$O \textsc{iv}$]$ & 25.8903 & $1.091 \pm 0.063$ & $3.898 \pm 0.092$\\
\hline
\end{tabular}
\begin{tablenotes}
\item\textbf{Notes.} $^a$Vacuum wavelengths from the compilation by \citet{vanhoof18}. $^b$Fluxes were summed over the pixels encompassed in the regions defined in Sect.\,\ref{subsect:mrs}. $^c$The SOR fluxes were scaled to the NOR regions as explained in Sect.\,\ref{subsect:mrs}.
\end{tablenotes}
\end{table}

The [N\,\textsc{ii}] $\lambda$\,5755 line is broad and significantly blue-shifted in the NOR and red-shifted in the SOR. The large flux ratio of this line compared to the [N\,\textsc{ii}] $\lambda$\,6583 indicates that there is leakage of emission from other regions, and the most likely is the shock emission from the ER where the [N\,\textsc{ii}] $\lambda$\,5755 line is stronger. The [O\,\textsc{i}] $\lambda\lambda$\,6300, 6364 lines are heavily contaminated by artefacts and their line fluxes cannot be measures accurately.
Except for [N\,\textsc{ii}], He\,\textsc{i}, [Fe\,\textsc{ii}], and [Ca\,\textsc{ii}], all lines are stronger in the SOR than in the NOR. The continuum is larger in the NOR than in the SOR but is not responsible for the difference in the strengths of the lines. 
In general, most lines appear to be brighter and more red-shifted in the NOR than in the SOR. The limited spectral resolution of MUSE makes it difficult to draw firm conclusions about differences in peak velocities (on the order of 10-20\,km\,s$^{-1}$) between different lines. The H$\beta$ (and to a lesser extent the H$\alpha$) line is contaminated by the surrounding medium and shows a behaviour opposite to most other lines. 

In Fig.\,\ref{fig:MUSE_lambda_and_fwhm}, we present a colourmap in the ORs of the velocity at maximum flux $V_\text{peak}$ of the [N\,\textsc{ii}] $\lambda$\,6583 line. In the regions contaminated by the ER, SN ejecta, or stars in the field of view, $V_\text{peak}$ takes extreme values, either because the line is blue- or red-shifted or because there is no line at all and only a flat continuum is contributing. These maps clearly highlight the regions where there is no contamination, as the rest of the ORs show a very homogeneous distribution of $V_\text{peak}$, well within the spectral resolution of the instrument. The same is observed for the other emission lines in the ORs presented in Fig.\,\ref{fig:images_MUSE}.

\section{JWST observations\label{sect:JWST}}
\subsection{JWST/NIRCam\label{subsect:nircam}}

\begin{figure}
\includegraphics[width=\linewidth]{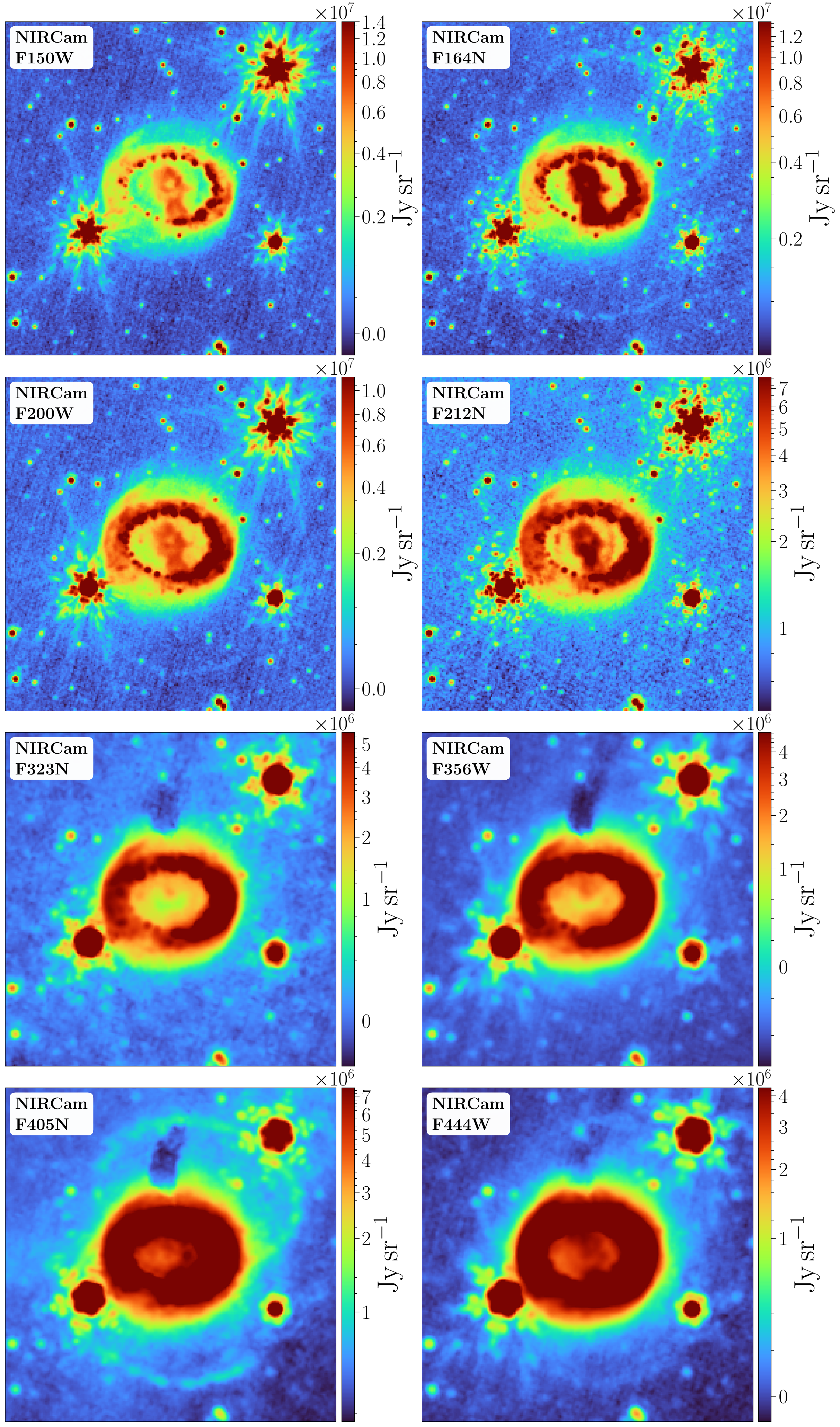}
\caption{JWST/NIRCam images of SN\,1987A in the different filters taken on September 1st and 2nd 2022 (days 12974 and 12975). The field of view for each image is $6\fds0 \times 6\fds0$. \label{fig:NIRCam_images}}
\end{figure}

The images in the different NIRCam filters are shown in Fig.\,\ref{fig:NIRCam_images}. The ORs are only clearly visible in the F164N, F200W, F405N, and F444W filters, to which the iron and hydrogen lines significantly contribute. Qualitatively, the brightest regions in the ORs seen in the NIRCam images are also the brightest in the HST images and MUSE data.

\begin{figure*}
\includegraphics[width=\linewidth]{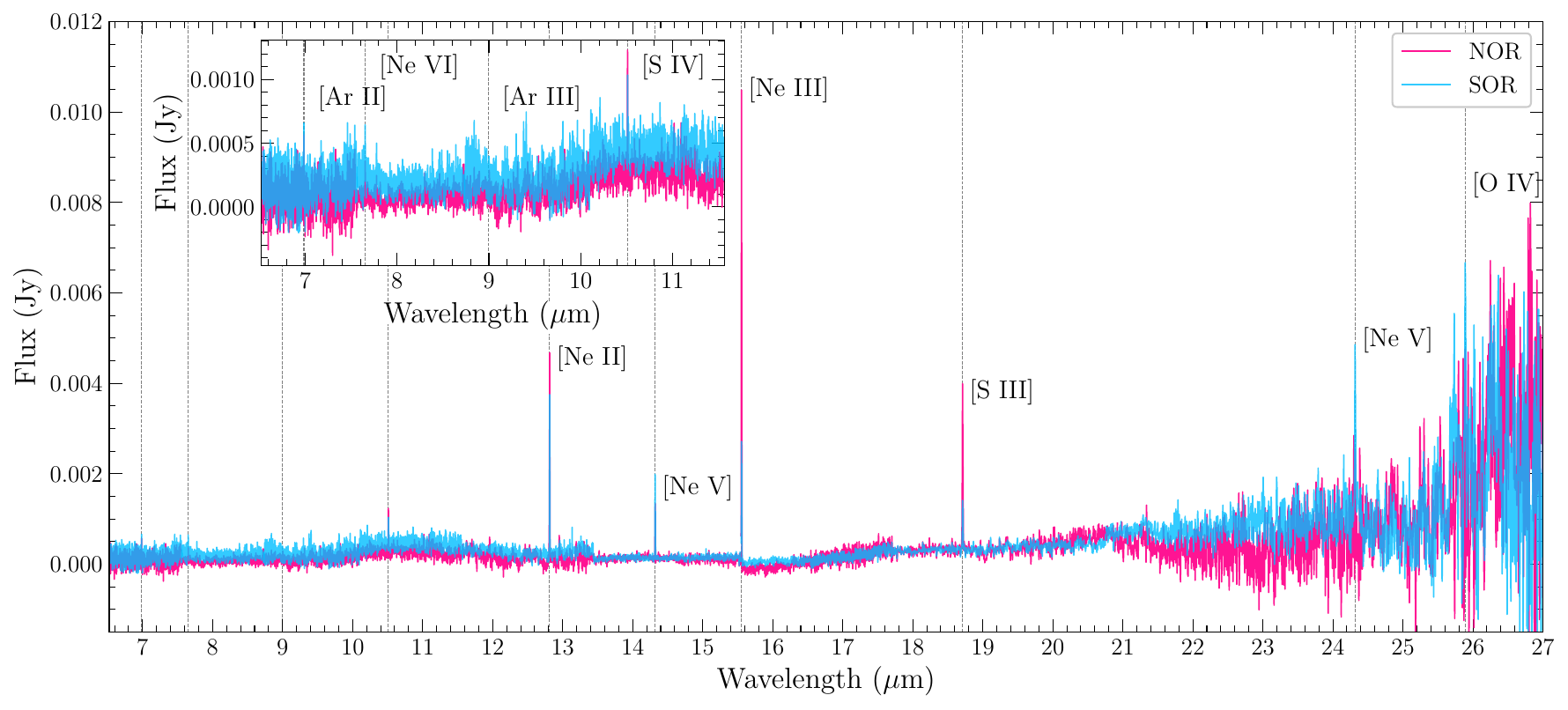}
\caption{JWST/MIRI/MRS mean background subtracted spectra of the NOR and SOR with main lines identified. \label{fig:JWST_spectra}}
\end{figure*}

To compute the fluxes in the ORs, we adopted the same regions as for the HST (Sect.\,\ref{subsect:lightcurves}) and MUSE (Sect.\,\ref{sect:MUSE}), except that we scaled them to account for the better spatial resolution of NIRCam. The fluxes were de-reddened adopting the grain model extinction law from \citet{weingartner01} as implemented in \citet{gordon24} for average LMC with a colour excess $E(B-V)=0.19$\,mag. Given that the number of pixels in the NOR and SOR are different (ratio of 1.39 in surface area for the F150W, F164N, F200W, and F212N filters, and 1.49 for the F323N, F356W, F405N, and F444W filters), we scaled the SOR by the ratio between the number of pixels in the NOR and SOR in the considered filter to allow for a quantitative comparison between both ORs. 

The flux measurements in the NOR and SOR in all NIRCam observations are provided in Table\,\ref{table:NIRCam_fluxes}, together with the main emission components contributing to these filters. We accounted for both statistical and systematic errors as for HST images (see Sect.\,\ref{subsect:lightcurves}). The fluxes are larger in the NOR than in the SOR, in all filters. The fluxes in the filters F212N and F323N are dominated by field stars and the flux in the filter F356W is dominated by the continuum.

\subsection{JWST/MIRI/MRS\label{subsect:mrs}}
We extracted background subtracted JWST/MIRI/MRS spectra for the NOR and SOR adopting regions that encompass as much of the ORs as possible without including the contaminated regions. Due to the broadening of the PSF with increasing wavelength channels, the exclusion regions were significantly larger at longer wavelengths. For channels 1L to 2L, one single background spectrum was extracted for both ORs, while for channels 3S to 4L, two dedicated background spectra could be extracted. We adopted background regions far from the edges of the cubes and where Cycles 1 and 2 overlap to maximise the S/N. We also avoided clear visible background structures.

The mean spectra and main line identifications are shown in Fig.\,\ref{fig:JWST_spectra}. The continuum in the SOR is slightly larger than that in the NOR  at all wavelengths.  Except for the [Ar\,\textsc{ii}] $\lambda$\,6.99, [Ne\,\textsc{vi}] $\lambda$\,7.65, [Ne\,\textsc{v}] $\lambda\lambda$\,14.32, 24.32, and [O\,\textsc{iv}] $\lambda$\,25.89 lines, all lines are stronger in the NOR than in the SOR.

\begin{figure}[t]
\includegraphics[width=\linewidth]{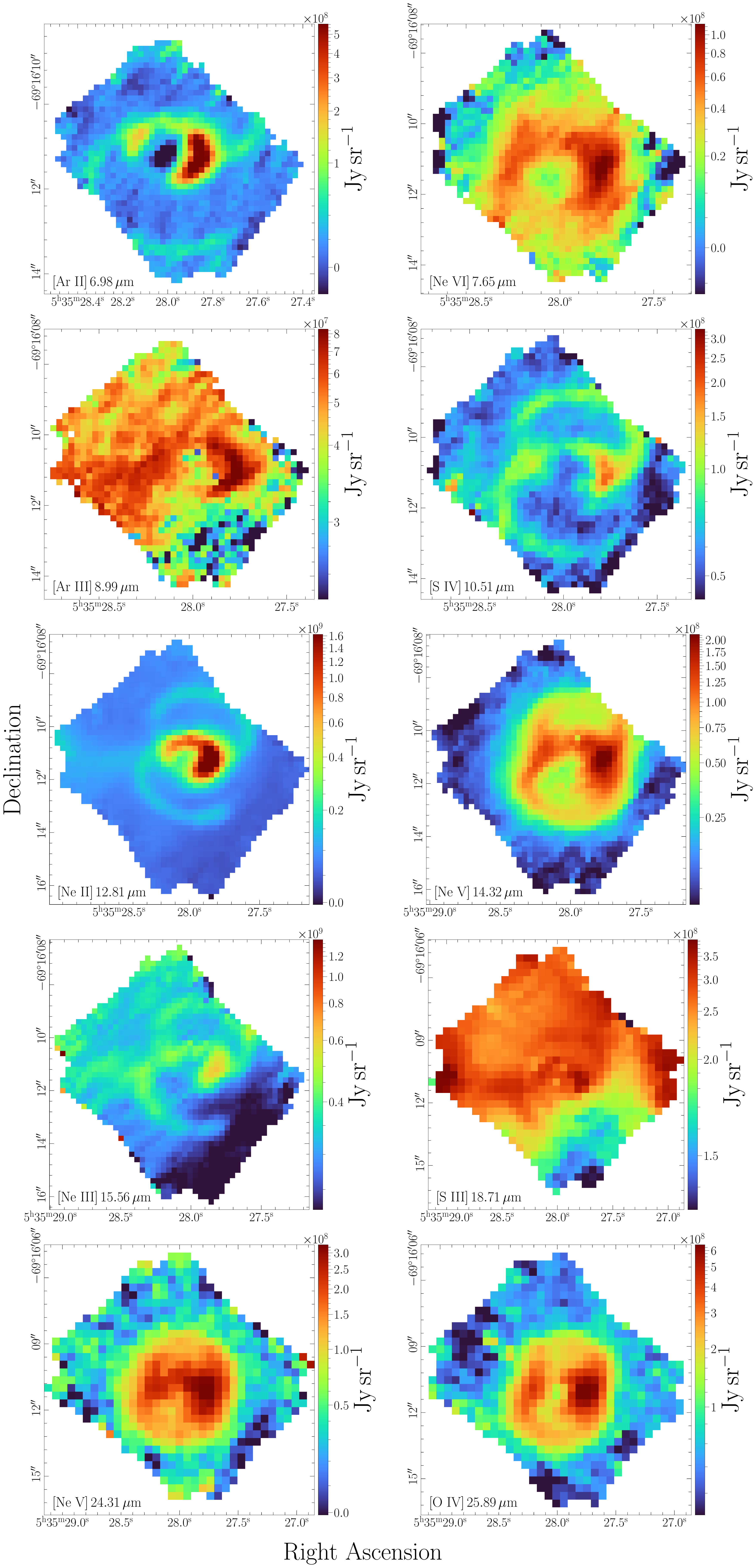}
\caption{JWST/MIRI/MRS slab images around the main emission components of the ORs (see Fig.\,\ref{fig:JWST_spectra}). From left to right and top to bottom: cubes 1L, 2S--M--L, 3S--M--L, 4S--M--L. The field of view is $3\fds2 \times 3\fds7$ for 1L, $4\fds0 \times 4\fds8$ for 2S--M--L, $5\fds2 \times 6\fds2$ for 3S--M--L, and $6\fds6 \times 7\fds7$ for 4S--M--L. \label{fig:JWST/MRS_images}}
\end{figure}

The slab images extracted around the main emission lines (denoted in Fig.\,\ref{fig:JWST_spectra}) are shown in Fig.\,\ref{fig:JWST/MRS_images}. Except for the last three cubes, the northest part of the NOR is never fully covered by the field of view of MIRI/MRS. In addition, in the first three images, parts of the SOR are not encompassed by the image field of view. The ORs are clearly distinguishable from the ER and inner SN ejecta in all cubes except for 2M, consistent with the lack of emission lines in the spectra at this wavelength.

We fitted the line-integrated flux of the emission lines in the NOR and SOR by summing over the pixels encompassed by the defined regions and removing the background contribution. The fluxes are constrained by the features outlined above. The fluxes were de-reddened adopting the same extinction curve as for NIRCam observations (Sect.\,\ref{subsect:nircam}). We scaled the SOR by the ratio between the number of pixels in the NOR and SOR to allow for a quantitative comparison between both ORs. The flux measurements in the NOR and SOR, in the MIRI/MRS emission lines are provided in Table\,\ref{table:MRS_fluxes}. We accounted for both statistical and systematic errors as for HST images (see Sect.\,\ref{subsect:lightcurves}).

\section{Discussion\label{sect:discussion}}
We discuss the optical light curve (Sect.\,\ref{subsect:diss_lightcurve}) and emission lines (Sect.\,\ref{subsect:diss_line_fluxes}) as indicators of temperature and electron density. A comparison between the ORs and ER in light of the latest studies of the ER is presented in Appendix \ref{sect:diss_ER}.

\subsection{Optical light curve\label{subsect:diss_lightcurve}}
\citet{tziamtzis11} presented light curves of the ORs from HST images in the F502N filter (see their figure 10). The light curves we present in Fig.\,\ref{fig:lightcurve} are smoother than theirs and do not show any large up-and-down variations between two consecutive epochs as we adopted larger apertures than they did, thus smoothing small pixel-to-pixel variations inherent to dim regions like the ORs. In particular, \citet{tziamtzis11} discarded the ACS observations taken at days 5795, 6122, and 6505 as they showed significant variations compared to previous epochs, which the authors attributed to reflections. We do not observe these effects in our light curve. If reflections would be present, they would be washed out by the large apertures adopted in this study.

The decay of the [O\,\textsc{iii}] (F502N filter) and H$\alpha$+[N\,\textsc{ii}] (F657N filter) emissions is a combination of recombination and cooling. We estimated the decay times $t_\text{decay}$ of these emissions by fitting the light curves in Sect.\,\ref{subsect:lightcurves} with the function 
\begin{equation}
f_\text{line} = a \exp\left({-\frac{t-t_\text{delay}}{t_\text{decay}}}\right)+c,
\end{equation}
where $t$ is the number of days since explosion, $t_\text{delay}$ is the light travel time estimated to be 600-1000\,days for the NOR and 100-600\,days for the SOR \citep{tziamtzis11}, and $a$ and $c$ are free parameters of the fit. For [O\,\textsc{iii}], $t_\text{decay}$ is 900 and 630-730 days for the NOR and SOR, respectively. 
For H$\alpha$+[N\,\textsc{ii}], $t_\text{decay}$ is 15\,870 and 7160 days for the NOR and SOR, respectively. As the flux ratio between [N\,\textsc{ii}] $\lambda$\,6548+$\lambda$\,6583 and H$\alpha$ decreased from 1.74 (NOR) and 2.24 (SOR) at day 5703, to 1.61 (NOR) and 1.94 (SOR) at day 8000 \citep[from UVES data of][]{tziamtzis11}, and to 1.32 (NOR) and 1.00 (SOR) at day 13182 (our MUSE data, see Table\,\ref{table:MUSE}), it affects the determination of individual H$\alpha$ and [N\,\textsc{ii}] decay times. 

\subsection{Emission line fluxes in the optical\label{subsect:diss_line_fluxes}}
\begin{figure*}
\includegraphics[width=\linewidth]{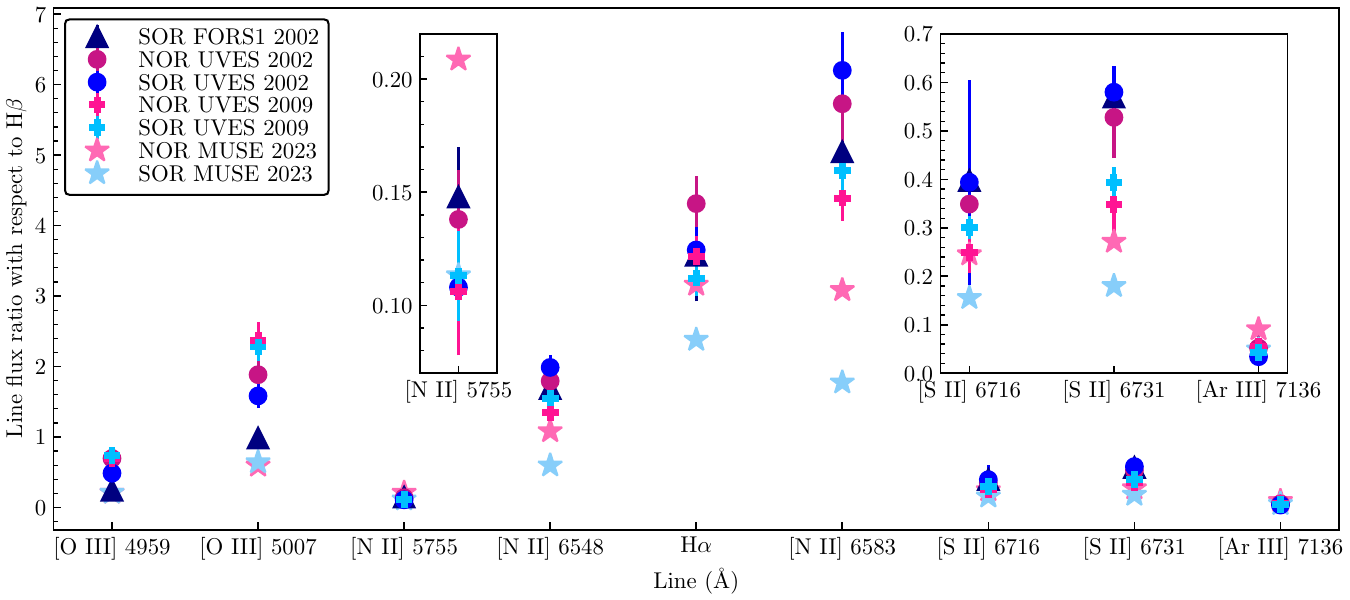}
\caption{Relative fluxes of lines with respect to H$\beta$. MUSE measurements are given in Sect.\,\ref{sect:MUSE}, while FORS1 and UVES measurements are from \citet{tziamtzis11}. \label{fig:line_emission_tziamtzis}}
\end{figure*}

\citet{tziamtzis11} presented emission line fluxes in small regions in the NOR and SOR from VLT/UVES data taken in October 2002 and January 2009 and in the SOR only from the VLT/FORS1 in December 2002 (see their tables 4, 6, and 7). A direct comparison between our MUSE emission line fluxes (see Sect.\,\ref{sect:MUSE}) and \citet{tziamtzis11} results is difficult because the regions adopted are different. Nonetheless, we compared the relative fluxes with respect to H$\beta$ assuming the small regions adopted in \citet{tziamtzis11} are representative of the NOR and SOR as a whole which, given the homogeneity of the ORs (in terms of spectra taken at different locations in the ORs despite the fact that the ORs consist of a number of discrete blobs), is reasonable (see Fig.\,\ref{fig:line_emission_tziamtzis}). As discussed in \citet{tziamtzis11}, the line ratios in the FORS1 2002 data are less accurate than those in the UVES data taken at the same epoch because of the much smaller aperture in FORS1. The [O\,\textsc{iii}] $\lambda\lambda$\,4959, 5007, [N\,\textsc{ii}] $\lambda\lambda$\,6548, 6583, H$\alpha$, and [S\,\textsc{ii}] $\lambda\lambda$\,6716, 6731  line fluxes have decreased with respect to H$\beta$ in both ORs over the 21-years period covered by the observations. Only the [Ar\,\textsc{iii}] $\lambda$\,7136 line flux has slightly increased over this period. The [N\,\textsc{ii}] $\lambda$\,5755 line flux ratio in the SOR has stayed constant, while in the NOR, it first decreased between years 2002 and 2009, then increased between years 2009 and 2023. As for the ER, the ORs are likely to consist of regions with a range of densities. Those regions with lower density recombine slower, which means that the [Ar\,\textsc{iii}]\,$\lambda$\,7136 line always come from regions with lower density than low-ionisation lines, including the Balmer lines.

An interesting feature is the line ratio of the [N\,\textsc{ii}] $\lambda\lambda$\,6548, 6583 doublet to H$\alpha$ in the ORs. The line ratio were $\sim$$2.0$$-$$2.4$ in the ORs between years 1994 and 2003 \citep{tziamtzis11}, but dropped in 2023 to 1.32 and 1.00 in the NOR and SOR, respectively, as measured from our MUSE observations. Comparison of our Fig.\,\ref{fig:spectra_MUSE} (middle plot) and figure 9 in \citet{tziamtzis11} shows that the [N\,\textsc{ii}] $\lambda$\,6583 line was much stronger than H$\alpha$ in the FORS1 2002 data, while it is now of similar strength as H$\alpha$ in the MUSE 2023 data, which explains the decrease in the line ratio between the [N\,\textsc{ii}] doublet and H$\alpha$.

\begin{figure}
\includegraphics[width=\linewidth]{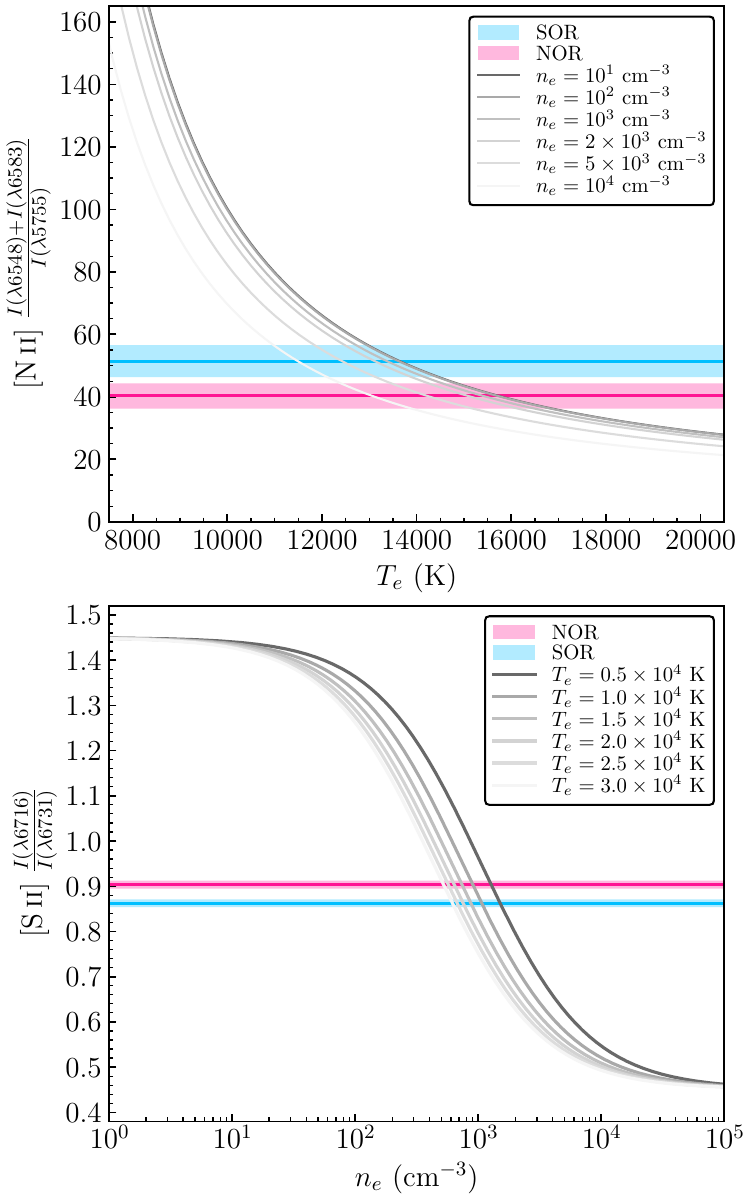}
\caption{\textit{Top panel:} [N\,\textsc{ii}] line ratio from Eq.\,\eqref{eqn:nii} as a function of temperatures $T_e$ for different electron densities $n_e$. \textit{Bottom panel:} [S\,\textsc{ii}] line ratio from Eq.\,\eqref{eqn:sii} as a function of $n_e$ for different $T_e$. Values for the NOR and SOR are indicated in both panels. \label{fig:temperature_density}}
\end{figure}

We made a joint estimation of the temperatures $T_e$ and electron densities $n_e$ from the [N\,\textsc{ii}] $\lambda\lambda$ 5575, 6548, 6583 lines and the [S\,\textsc{ii}] $\lambda\lambda$ 6716, 6731 lines.

We estimated the temperature $T_e$ from the analytic expressions for the lines ratios of \citet{osterbrock06}:
\begin{equation}
\label{eqn:nii}
\frac{I(\lambda\,6548)+I(\lambda\,6583)}{I(\lambda\,5755)} = \frac{8.23\exp{(2.50\times 10^4 T_e^{-1})}}{1+4.4\times 10^{-3} n_e T_e^{-0.5}}.
\end{equation}
Given that $n_e$ cannot be estimated based on [N\,\textsc{ii}] lines directly, we assumed a conservative value between 10 and 5000\,cm$^{-3}$. The upper limit is based on the argument that the highest electron density in the N\,\textsc{ii} region is not higher than the highest value from the S\,\textsc{ii} region (see below). Given that the 
[N\,\textsc{ii}] $\lambda$\,5755 line is heavily contaminated by the surrounding medium (Sect.\,\ref{sect:MUSE}), we extracted small regions in the northeast of the NOR and in the southwest of the SOR. The [N\,\textsc{ii}] $\lambda$\,5755 line is very faint and its flux hard to estimate. We obtained ratios of $40\pm4$ and $51\pm5$ for the NOR and the SOR, respectively. It corresponds to temperatures of 13400-16900\,K and 11800-14500\,K for the NOR and SOR, respectively, as shown in Fig.\,\ref{fig:temperature_density}. 
For the NOR, this $T_e$ is slightly larger than the value from the [N\,\textsc{ii}] lines of 12000\,K derived from the UVES data \citep[days 5704-5705 and 7944-8021,][]{tziamtzis11}.
For the SOR, this $T_e$ is in agreement with the values from the [N\,\textsc{ii}] lines of 13000-14000\,K, 10000-11000\,K and 11500-12500\,K derived from the FORS1 (day 5791), UVES (days 5704-5705), and UVES (days 7944-8021) data, respectively \citep{tziamtzis11}. 

We estimated the electron density using the analytic model of a five-level atom of \citet{castaneda92}:
\begin{equation}
\label{eqn:sii}
\frac{I(\lambda\,6716)}{I(\lambda\,6731)} = 1.45\frac{1+4.18\times10^{-6}n_eT_e^{1/2}}{1+13.42\times10^{-6}n_eT_e^{1/2}}.
\end{equation}
Given that $T_e$ cannot be estimated based on [S\,\textsc{ii}] lines, we assumed a conservative value between 10000 and 20000\,K. [S\,\textsc{ii}] is likely to come from gas with lower temperature (on average) than [N\,\textsc{ii}], hence the upper limit on $T_e$. We obtained $n_e = 610-670$\,cm$^{-3}$ for the NOR and $n_e = 720-790$\,cm$^{-3}$ for the SOR, as shown in Fig.\,\ref{fig:temperature_density}.
These values are a factor 3-4 lower than the upper limit derived by \citet{tziamtzis11}.

We computed line ratios between different small regions in the ORs and do not see significant differences indicating varying density and/or temperature across the ORs.

\section{Conclusion\label{sect:conclusion}}
We analysed HST observations of the ORs of SN\,1987A in the F502N filter covering the [O\,\textsc{iii}] $\lambda$\,5007 line from year 1994 to 2022 and in the F657N covering the H$\alpha$ line from 2009 to 2022 (see Sect.\,\ref{sect:HST}), MUSE observations taken in  2023 in the narrow-field mode (see Sect.\,\ref{sect:MUSE}), JWST/NIRCam observations taken in 2022 in the F150W, F164N, F200W, F212N, F323N, F356W, F405N, and F444W filters (see Sect.\,\ref{subsect:nircam}), and JWST/MIRI/MRS IFU observations taken in 2022 and 2024 (see Sect.\,\ref{subsect:mrs}).

The NOR is $\sim$$10$\% brighter than the SOR in all HST and JWST/NIRCam observations, partly due to the larger continuum contribution. All emission lines are stronger in the SOR than in the NOR except for the [N\,\textsc{ii}], He\,\textsc{i}, [Fe\,\textsc{ii}], and [Ca\,\textsc{ii}] optical lines, and the [Ar\,\textsc{ii}], [Ne\,\textsc{vi}], [Ne\,\textsc{v}], and [O\,\textsc{iv}] mid-infrared lines.

The HST observations show a decrease in flux in the ORs in both filters. This decrease is more pronounced from days 2537 to 3792 than afterwards. At later times, from days 8329 to 12\,980, it is more pronounced in the F657N filter than in the F502N filter. Both western and eastern parts of the ORs decrease at the same rate. The western part of the NOR is, however, systematically brighter than its eastern part while the eastern part of the SOR is systematically brighter than its western part. We estimated the decay times for [O\,\textsc{iii}] to be 900 and 630-730\,days for the NOR and SOR, respectively, and for H$\alpha$+[N\,\textsc{ii}] to be 15\,870 and 7160\,days for the NOR and SOR, respectively.  

A comparison with \citet{tziamtzis11} showed that all optical line fluxes have decreased while the [Ar\,\textsc{iii}] $\lambda$\,7136 line flux has increased with respect to H$\beta$ in both ORs over the 21-years period covered by the observations. We conclude that the [Ar\,\textsc{iii}]\,$\lambda$ 7136 line always comes from regions with lower density than low-ionisation lines as these regions recombine slower. Furthermore, the line ratio of the [N\,\textsc{ii}] $\lambda\lambda$\,6548, 6583 doublet to H$\alpha$ has decreased from $\sim$$2$ in 1994-2003 (days 2537-6122) to $\sim$$1$ in 2023 (day 13182). It is caused by the [N\,\textsc{ii}] $\lambda$\,6583 being much stronger than H$\alpha$ in the FORS1 2002 data (day 5789), but now of similar strength as H$\alpha$ in the MUSE 2023 data.

Finally, the spectra of the ORs significantly differ from the ER' spectrum, not only in the lines detected, but also in the line ratios of given lines.

From these investigations, we conclude that the SN ejecta have not interacted with the ORs yet. Indeed, the line width of the SN ejecta is thousands of \,km\,s$^{-1}$ \citep{larsson23}, much larger than the line width of the ORs that are smaller than the spectral resolution of MUSE and JWST/MIRI/MRS. \citet{larsson23} and \citet{matsuura24} showed on JWST/NIRSpec and NIRCam data, respectively, that the SN ejecta passed the ER and started interacting with the medium near the ER. The finding that all lines decrease in strength means that reonisation by the X-ray emission from the SN ejecta-ER collisions did not occur.

The ORs will likely keep on fading until the SN ejecta collide with and heat them. According to \citet{tziamtzis11}, the SN ejecta blast wave could reach the ORs in about 5-10 years from now, with considerable uncertainties in this estimate. It may not be until very much later that this happens (judging from MUSE images, it could be closer to 2040 than 2030). The interaction between the SN ejecta and the ORs will be less spectacular than the interaction between the SN ejecta and the ER when the SN ejecta reached the ER in 1995 because the SN ejecta have expanded and decreased in density. Continued monitoring of SN\,1987A and its ring system from X-ray to MIR is essential to capture this instant when the SN ejecta reach the ORs. It still remains to be seen if and how the ER, NOR, and SOR are physically connected or if they are physically separated and emerged from different past events of the progenitor's life. It is also crucial to perform 2D and 3D hydrodynamical simulations for various density distributions in order to investigate when and how the SN ejecta will plough out through the ORs.

\begin{acknowledgement}
The authors thank the anonymous referee for their constructive report. S. Rosu is supported by the Swiss National Funding under project No 212143. P. J. Kavanagh acknowledges support from the Research Ireland Pathway programme under Grant Number 21/PATH-S/9360. R. D. Gehrz was supported, in part, by the United States Air Force. M. Meixner acknowledges support through NASA/JWST Task plan 71-209636. M. Meixner acknowledges that a portion of her research was carried out at the Jet Propulsion Laboratory, California Institute of Technology, under a contract with the National Aeronautics and Space Administration (80NM0018D0004). This research is based in part on observations made with ESO Telescopes at the La Silla Paranal Observatory under programme ID 11.25AR.001. This research is based in part on observations made with the NASA/ESA Hubble Space Telescope obtained from the Space Telescope Science Institute, which is operated by the Association of Universities for Research in Astronomy, Inc., under NASA contract NAS 5–26555. These observations are associated with programs 5203, 6020, 6437, 8243, 9114, 9428, 9992, 10263, 10549, 10867, 11653, 11973, 12241, 13405, 14333, 14753, 15256, 15503, 15928, 16265, 16789. This work is based in part on observations made with the NASA/ESA/CSA James Webb Space Telescope. The data were obtained from the Mikulski Archive for Space Telescopes at the Space Telescope Science Institute, which is operated by the Association of Universities for Research in Astronomy, Inc., under NASA contract NAS 5-03127 for JWST. These observations are associated with program 1232, 1726, and 2763.
\end{acknowledgement}


\begin{appendix} 
\section{HST observations between years 1994 and 2022\label{appendix:HST_images}}
\begin{figure*}
\includegraphics[width=\linewidth]{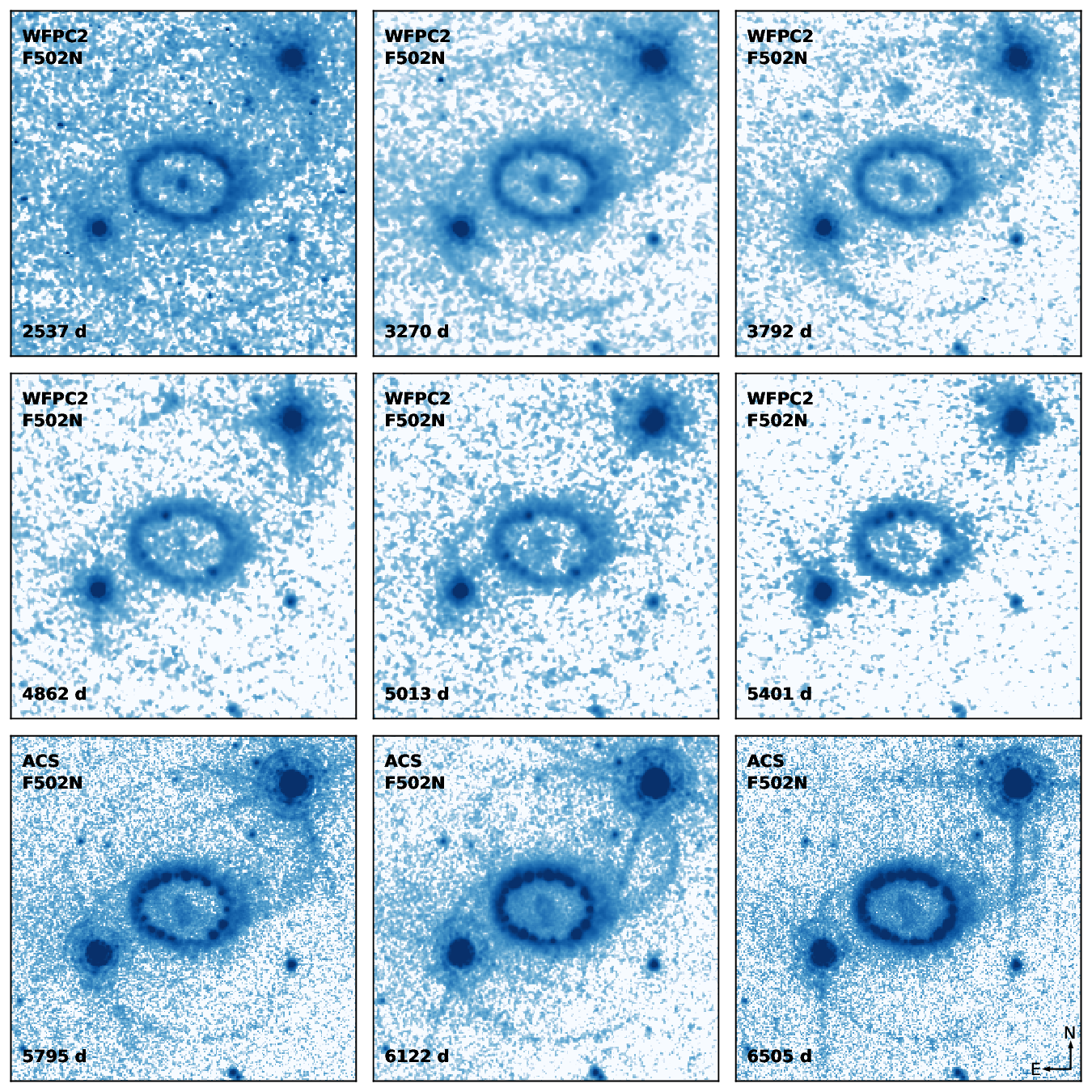}
\caption{HST/WFPC2 and ACS images of SN\,1987A taken in the F502N filter between 2537 and 6505 days after the explosion. The images were scaled by an asinh function and the color scales were chosen differently for the two instruments to highlight the weak emission in the ORs. The field of view for each image is $6\fds0 \times 6\fds0$.}
\label{fig:obs_F502N1}
\end{figure*}

\begin{figure*}
\includegraphics[width=\linewidth]{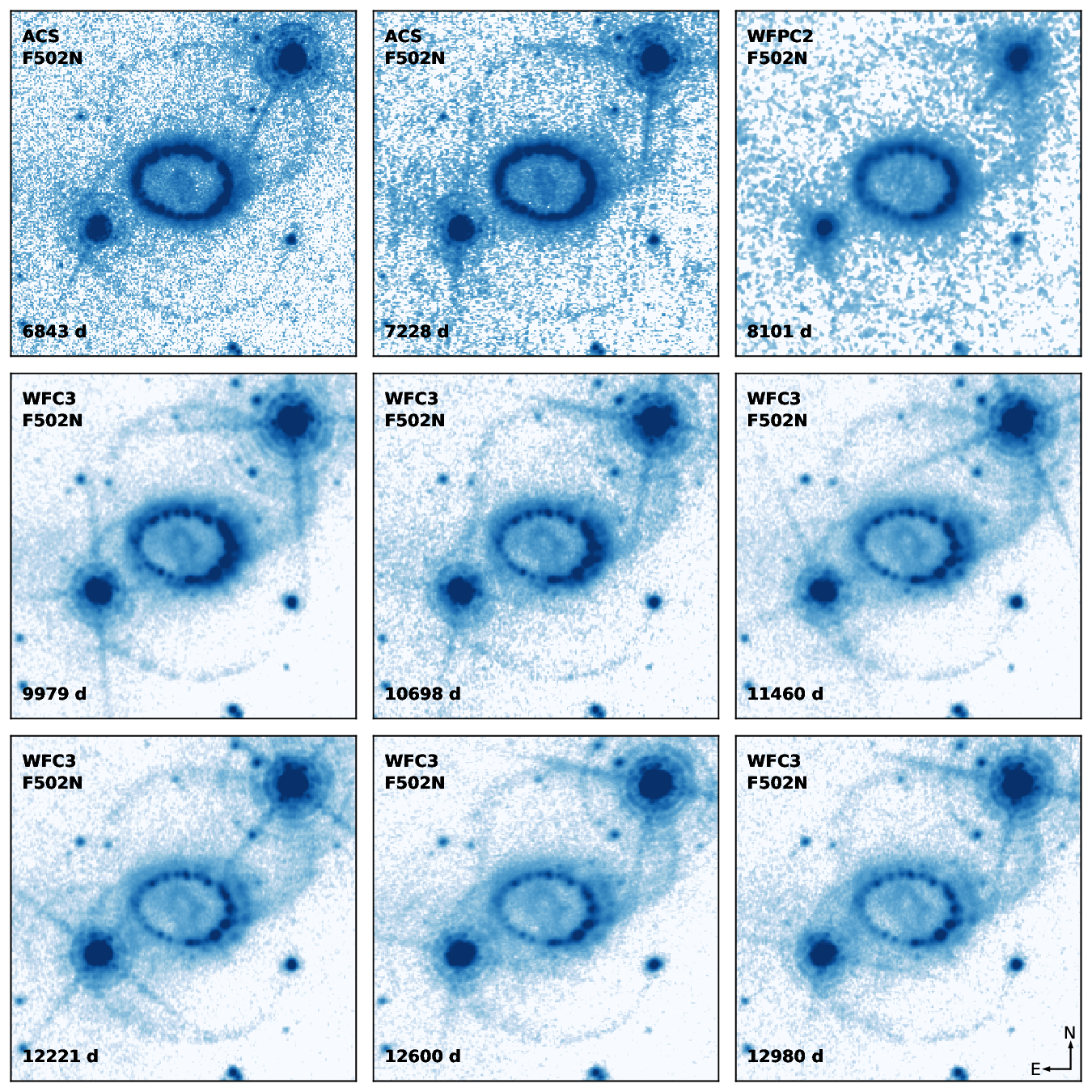}
\caption{HST/ACS, WFPC2, and WFC3 images of SN\,1987A taken in the F502N filter between 6843 and 12\,980 days after the explosion. The images were scaled by an asinh function and the color scales were chosen differently for the three instruments to highlight the weak emission in the ORs. The field of view for each image is $6\fds0 \times 6\fds0$.}
\label{fig:obs_F502N2}
\end{figure*}

\begin{figure*}
\includegraphics[width=\linewidth]{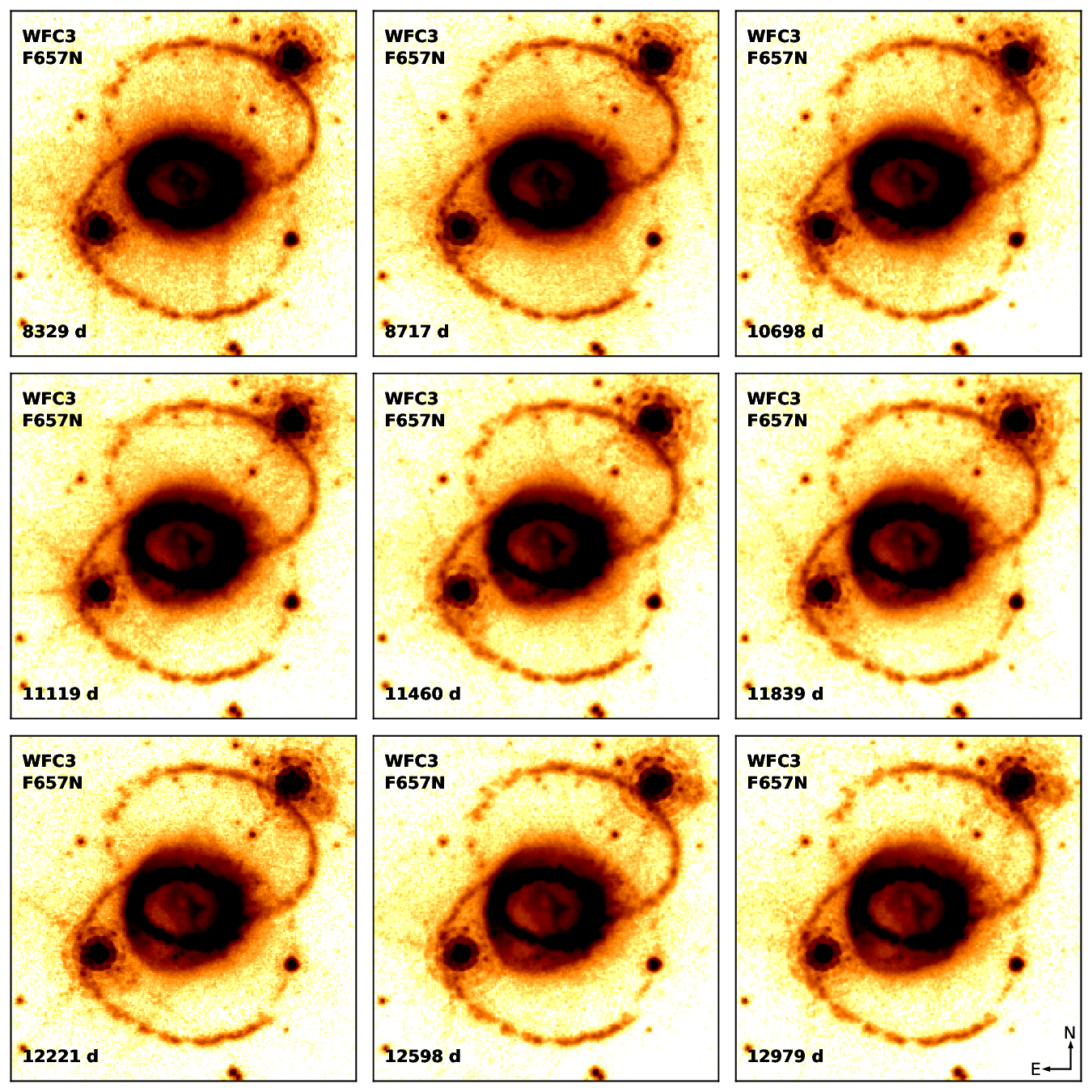}
\caption{HST/WFC3 images of SN\,1987A taken in the F657N filter between 8329 and 12\,979 days after the explosion. The images were scaled by an asinh function to highlight the weak emission in the ORs. The field of view for each image is $6\fds0 \times 6\fds0$.}
\label{fig:obs_F657N}
\end{figure*}

HST observations analysed in this paper are shown in Figs\,\ref{fig:obs_F502N1}, \ref{fig:obs_F502N2}, and \ref{fig:obs_F657N}). The ORs are more visible in the WFC3/F502N than in ACS/F502N and WFPC2/F502N images. It also highlight which images are affected by CTE losses.

\section{Diffraction spike contributions\label{appendix:diffraction_spikes}} 
The diffraction spikes contributions to the fluxes in the HST images (see Sect.\,\ref{subsect:lightcurves}) were removed following the same procedure as in \citet{rosu24}. We provide in Table\,\ref{table:diffraction_spikes} the reduction of the flux after removal of the diffraction spike contributions in the ORs in the different filters. 

\begin{table}[h]
\caption{Reduction of the flux (in \%) in the NOR and SOR regions after removal of the diffraction spike contributions from Stars 2 and 3 in the F502N and F657N filters in the HST images between Epochs 2537 and 12\,980.}
\centering
\begin{tabular}{llllll}
\hline\hline
Epoch & Filter & NOR & Star & SOR & Star \\
\hline
~~2537 & F502N & 0.15 & 3 & 5.49 & 2 and 3\\
~~3270 & F502N & 0.42 & 3 & 2.56 & 2 and 3\\
~~3792 & F502N & 1.26 & 2 and 3 & 2.85 & 3 \\
~~5795 & F502N & 9.09 & 2 and 3 & 12.06 & 3\\
~~6122 & F502N & 1.16 & 2 and 3 & 2.72 & 2 and 3 \\
~~6505 & F502N & 7.76 & 2 and 3 & 0.18 & 2 \\
~~6843 & F502N & $-$0.34 & 3 & 5.42 & 2 and 3\\
~~7228 & F502N & 14.43 & 2 and 3 & $-$2.94 & 2 \\
~~9979 & F502N & 10.43 & 2 & 0.41 & 2 and 3 \\
10\,698 & F502N & 16.87 & 2 and 3 & 6.07 & 2 and 3 \\
11\,460 & F502N & 5.86 & 2 and 3 & 14.32 & 3\\
12\,221 & F502N & $-$0.28 & 3 & 11.87 & 2 and 3\\
12\,600 & F502N & 14.22 & 2 and 3 & 0.96 & 2 and 3 \\
12\,980 & F502N & 15.32 & 2 and 3 & 1.86 & 2 and 3\\
~~8329 & F657N & 2.50 & 2 and 3 & 0.43 & 2 \\
~~8717 & F657N & 1.47 & 2 and 3 & 1.31 & 3\\
10\,698 & F657N & 3.91 & 2 and 3 & 0.70 & 2 and 3\\
11\,119 & F657N & 0 & ... & 1.31 & 3\\
11\,460 & F657N & 1.09 & 2 and 3 & 1.43 & 3\\
11\,839 & F657N & 0 & ... & 1.39 & 3\\
12\,221 & F657N & $-$0.02 & 3 & 1.56 & 2 and 3 \\
12\,598 & F657N & 3.14 & 2 and 3 & 0.44 & 2 and 3 \\
12\,979 & F657N & 3.74 & 2 and 3 & 0.46 & 2 and 3\\
\hline
\end{tabular}
\label{table:diffraction_spikes}
\end{table}

\section{Comparison with the ER\label{sect:diss_ER}}
\citet{kangas22} presented HST/Space Telescope Imaging Spectrograph (STIS) spectra of the ER from 2017 \citep[for detailed line identifications see][]{rosu24}. All optical lines seen in the ORs are present in the ER. However, their respective line fluxes are significantly different. While the [O\,\textsc{iii}] $\lambda\lambda$\,4959, 5007 line fluxes are of the same order of magnitude as H$\beta$ in the ORs, they are significantly lower than H$\beta$'s in the ER. The [N\,\textsc{ii}] $\lambda\lambda$\,5755, 6548, 6583 lines fluxes are significantly lower than H$\beta$ in the ER while they are of the same order of magnitude in the ORs. The [S\,\textsc{ii}] $\lambda\lambda$\,6716, 6731 lines are only barely visible in the ER. Furthermore, several lines prominent in the ER are not observed in the ORs, including He\,\textsc{ii}, [O\,\textsc{i}], [N\,\textsc{i}], and iron lines ([Fe\,\textsc{vii}], [Fe\,\textsc{x}], [Fe\,\textsc{xi}], and [Fe\,\textsc{xiv}]). 

\begin{figure}
\includegraphics[width=\linewidth]{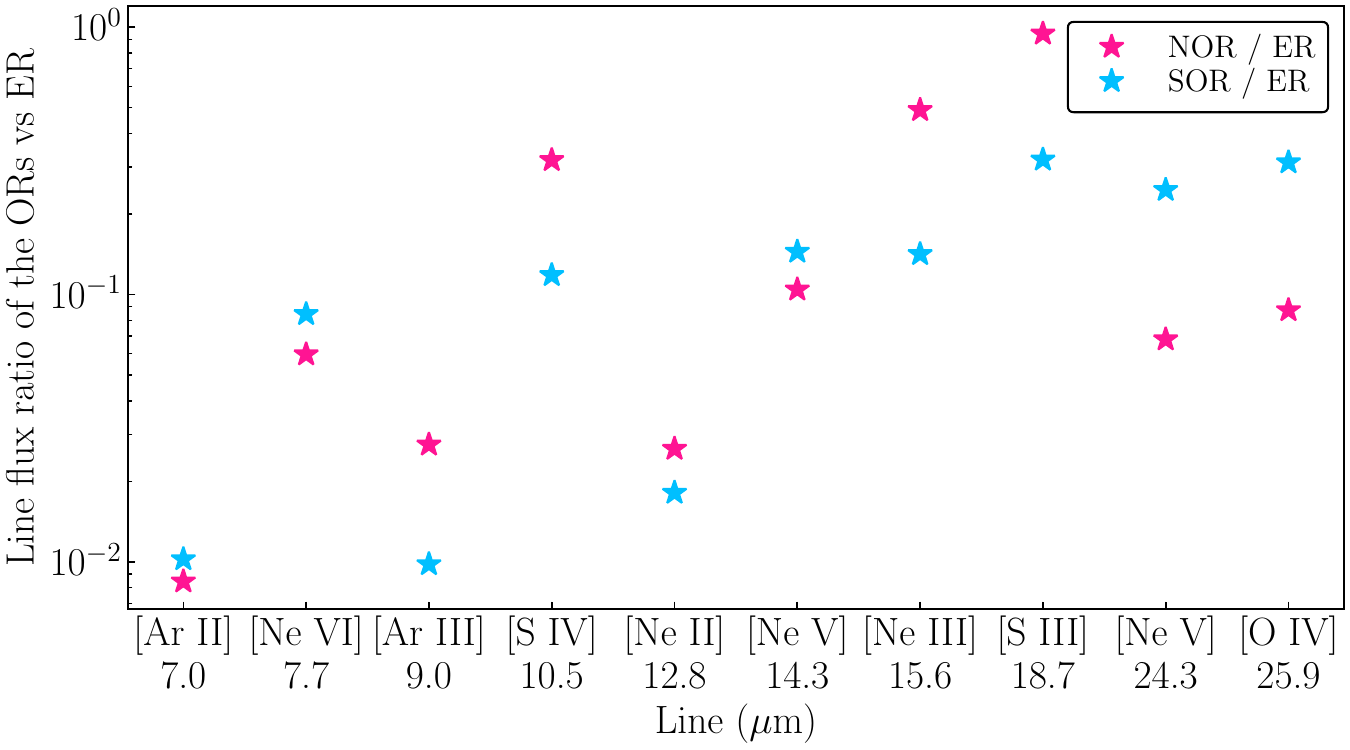}
\caption{Line flux ratio of the ORs over the ER in the JWST/MIRI/MRS data. The fluxes from the ER are taken from \citet{jones23b}. \label{fig:ER_ORs}}
\end{figure}

\citet{jones23b} presented MRS spectra of the ER on day 12\,927. All lines seen in the ORs are present in the ER. However, their respective line fluxes are significantly different. Figure \ref{fig:ER_ORs} shows the ratio between NOR or SOR line fluxes and ER line fluxes. While the exact values do not mean much, as the apertures adopted for the ORs and ER are not comparable, it is interesting to see the differences of three orders of magnitude between individual line flux ratios. In addition, lines of nitrogen, sodium, hydrogen, helium, and iron are observed in the ER \citep{jones23b, kavanagh25}, but not in the ORs.

These comparisons demonstrate, if necessary, that it is crucial to consider the ORs as different than the ER, as their spectra significantly differ, notably bacause the ER and ORs differ in initial level of ionisation after the UV flash, but also because the ER is ionised by shock interaction with the SN ejecta.

\end{appendix}

\begin{thebibliography}{99}
\bibitem[Alp et al.(2018)]{alp18}
Alp, D., Larsson, J., Fransson, C., et al.\ 2018, \apj, 864, 174
\bibitem[Argyriou et al.(2023)]{argyriou23}
Argyriou, I., Glasse, A., Law, D. R., et al.\ 2023, \aap, 675, A111
\bibitem[Arnett et al.(1989)]{arnett89}
Arnett, W. D., Bahcall, J. N., Kirshner, R. P., \& Woosley, S. E.\ 1989, \araa, 27, 629
\bibitem[Bacon et al.(2010)]{bacon10}
Bacon, R., Accardo, M., Adjali, L., et al.\ 2010, Proceedings of the SPIE, Volume 7735, id. 773508
\bibitem[Castañeda et al.(1992)]{castaneda92}
Castañeda, H. O., Vílchez, J. M., \& y Copetti, M. V. F.\ 1992, A\&A 260, 370
\bibitem[Crotts \& Heathcote(2000)]{crotts00}
Crotts, A. P. S., \& Heathcote, S. R.\ 2000, \apj, 528, 426
\bibitem[France et al.(2011)]{france11}
France, K., McCray, R., Penton, S. V., et al.\ 2011, \apj, 743, 186
\bibitem[Fransson et al.(2013)]{fransson13}
Fransson, C., Larsson, J., Spyromilio, J., et al.\ 2013, \apj, 768, 88
\bibitem[Fruchter et al.(2010)]{fruchter10}
Fruchter, A. S., Hack, W., Dencheva, N., et al.\ 2010, STSCI Calibration Workshop Proceedings, Baltimore, MD, 21-23 July 2010, eds. S. Deustua \& C. Oliveira, STSCI, pp 376-381
\bibitem[Gordon(2024)]{gordon24}
Gordon, K. D.\ 2024, Journal of Open Source Software, 9, 100, 7023
\bibitem[Gröningsson et al.(2008)]{groningsson08}
Gröningsson, P., Fransson, C., Lundqvist, P., et al.\ 2008, \aap, 479, 761
\bibitem[Hoffmann et al.(2021)]{hoffmann21}
Hoffmann, S. L., Mack, J., Avila, R., et al.\ 2021, The DrizzlePac Handbook (Baltimore: STScI)
\bibitem[Jones et al.(2023a)]{jones23a}
Jones, O. C., \'Alvarez-M\'arquez, J., Sloan, G. C., et al.\ 2023a, \mnras, 523, 2519
\bibitem[Jones et al.(2023b)]{jones23b}
Jones, O. C., Kavanagh, P. J., Barlow, M. J., et al.\ 2023b, \apj, 958, 95
\bibitem[Kangas et al.(2022)]{kangas22}
Kangas, T., Fransson, C., Larsson, J., et al.\ 2022, \mnras, 511, 2977
\bibitem[Kavanagh et al.(2025)]{kavanagh25}
Kavanagh, P. J., Barlow, M. J., Fransson, C., et al. 2025, \apj, accepted [arXiv: 2604.09211]
\bibitem[Kunkel et al.(1987)]{kunkel87}
Kunkel, W., Madore, B., Shelton, I., et al.\ 1987, IAU Circ., 4316, 1
\bibitem[Larsson et al.(2019a)]{larsson19a}
Larsson, J., Fransson, C., Alp, D., et al.\ 2019a, \apj, 886, 147
\bibitem[Larsson et al.(2023)]{larsson23}
Larsson, J., Fransson, C., Sargent, B. et al.\ 2023, \apjl, 949, L27
\bibitem[Larsson et al.(2025)]{larsson25}
Larsson, J., Fransson, C., Kavanagh, P. J., et al.\ 2025, \apj, 991, 130
\bibitem[Maíz Apellániz et al.(2014)]{maiz14}
Maíz Apellániz, J., Evans, C. J., Barbá, R. H., et al.\ 2014, \aap, 564, A63
\bibitem[Matsuura et al.(2024)]{matsuura24}
Matsuura, M., Boyer, M., Arendt, R. G., et al.\ 2024, \mnras, 532, 3625
\bibitem[McCray(1993)]{mccray93}
McCray, R. 1993, \araa, 31, 175
\bibitem[McCray \& Fransson(2016)]{mccray16}
McCray, R., \& Fransson, C.\ 2016, \araa, 54, 19
\bibitem[Morris \& Podsiadlowski(2007)]{morris07}
Morris, T., \& Podsiadlowski, P.\ 2007, Science, 315, 1103
\bibitem[Morris \& Podsiadlowski(2009)]{morris09}
Morris, T., \& Podsiadlowski, P.\ 2009, \mnras, 399, 515
\bibitem[Osterbrock \& Ferland(2006)]{osterbrock06}
Osterbrock, D. E., \& Ferland, G. J.\ 2006, Astrophysics of Gaseous Nebulae and Active Galactic Nuclei, Second Edition, University Science Books, Sausalito, California
\bibitem[Panagia et al.(1991)]{panagia91}
Panagia, N., Gilmozzi, R., Macchetto, F., Adorf, H.-M., \& Kirshner, R. P.\ 1991, \apjl, 380, L23
\bibitem[Pietrzyński et al.(2019)]{pietrzynski19}
Pietrzyński, G., Graczyk, D., Gallenne, A., et al.\ 2019, Nature, 567, 200
\bibitem[Plait et al.(1995)]{plait95}
Plait, P. C., Lundqvist, P., Chevalier, R. A., \& Kirshner, R. P.\ 1995, \apj, 439, 730
\bibitem[Rosu et al.(2024)]{rosu24}
Rosu, S., Larsson, J., Fransson, C., et al.\ 2024, \apj, 966, 238
\bibitem[Soto et al.(2016)]{soto16}
Soto, K.~T., Lilly, S.~J., Bacon, R., et al.\ 2016, \mnras, 458, 3210
\bibitem[Sugerman et al.(2005)]{sugerman05}
Sugerman, B. E. K., Crotts, A. P. S., Kunkel, W. E., Heathcote, S. R., \& Lawrence, S. S.\ 2005, \apjs, 159, 60
\bibitem[Tziamtzis et al.(2011)]{tziamtzis11}
Tziamtzis, A., Lundqvist, P., Gr\"oningsson, P., \& Nasoudi-Shoar, S.\ 2011, \aap, 527, A35
\bibitem[Van Hoof(2018)]{vanhoof18}
Van Hoof, P. A. M. 2018, Galaxies, 6, 63,
\bibitem[Weilbacher et al.(2020)]{weilbacher20}
Weilbacher, P. M., Palsa, R., Streicher, O., et al.\ 2020, \aap, 641, A28
\bibitem[Weingartner \& Draine(2001)]{weingartner01}
Weingartner, J. C., \& Draine, B. T.\ 2001, \apj, 548, 296
\bibitem[Wells et al.(2015)]{wells15}
Wells, M., Pel, J. W., Glasse, A., et al.\ 2015, \pasp, 127, 646

\end{thebibliography}
\end{document}